\documentclass[twocolumn]{aastex631}

\shorttitle{Exploring the Fate of Stellar core collapse with SRNs}
\shortauthors{Ashida and Nakazato}
\graphicspath{{./}{figures/}}
\begin{document}

\title{Exploring the Fate of Stellar Core Collapse with Supernova Relic Neutrinos}

\author[0000-0003-4136-2086]{Yosuke Ashida}
\affiliation{Dept. of Physics and Wisconsin IceCube Particle Astrophysics Center, University of Wisconsin–Madison, Madison, WI 53706, USA}

\author[0000-0001-6330-1685]{Ken'ichiro Nakazato}
\affiliation{Faculty of Arts and Science, Kyushu University, Fukuoka 819-0395, Japan}

%\collaboration{6}{(AAS Journals Data Editors)}

%% Note that the \and command from previous versions of AASTeX is now
%% depreciated in this version as it is no longer necessary. AASTeX 
%% automatically takes care of all commas and "and"s between authors names.

%% AASTeX 6.31 has the new \collaboration and \nocollaboration commands to
%% provide the collaboration status of a group of authors. These commands 
%% can be used either before or after the list of corresponding authors. The
%% argument for \collaboration is the collaboration identifier. Authors are
%% encouraged to surround collaboration identifiers with ()s. The 
%% \nocollaboration command takes no argument and exists to indicate that
%% the nearby authors are not part of surrounding collaborations.

%% Mark off the abstract in the ``abstract'' environment. 
\begin{abstract}

Core collapse of massive stars leads to different fates for various physical factors, which gives different spectra of the emitted neutrinos. We focus on the supernova relic neutrinos (SRNs) as a probe to investigate the stellar collapse fate. We present the SRN fluxes and event rate spectra at a detector for three resultant states after stellar core collapse, the typical mass neutron star, the higher mass neutron star, or the failed supernova forming a black hole, based on different nuclear equations of state. Then possible SRN fluxes are formed as mixtures of the three components. We also show the expected sensitivities at the next-generation water-based Cherenkov detectors, SK-Gd and Hyper-Kamiokande, as constraining the mixture fractions. This study provides a practical example of extracting astrophysical constraints through SRN measurement. 

\end{abstract}

%% Keywords should appear after the \end{abstract} command. 
%% The AAS Journals now uses Unified Astronomy Thesaurus concepts:
%% https://astrothesaurus.org
%% You will be asked to selected these concepts during the submission process
%% but this old "keyword" functionality is maintained in case authors want
%% to include these concepts in their preprints.

%% From the front matter, we move on to the body of the paper.
%% Sections are demarcated by \section and \subsection, respectively.
%% Observe the use of the LaTeX \label
%% command after the \subsection to give a symbolic KEY to the
%% subsection for cross-referencing in a \ref command.
%% You can use LaTeX's \ref and \label commands to keep track of
%% cross-references to sections, equations, tables, and figures.
%% That way, if you change the order of any elements, LaTeX will
%% automatically renumber them.
%%
%% We recommend that authors also use the natbib \citep
%% and \citet commands to identify citations.  The citations are
%% tied to the reference list via symbolic KEYs. The KEY corresponds
%% to the KEY in the \bibitem in the reference list below. 

%%---------------------------------------------------------------------------------
\section{Introduction} \label{sec:intro}

\begin{figure}[htbp]
  \begin{center}
    \includegraphics[clip,width=8.6cm]{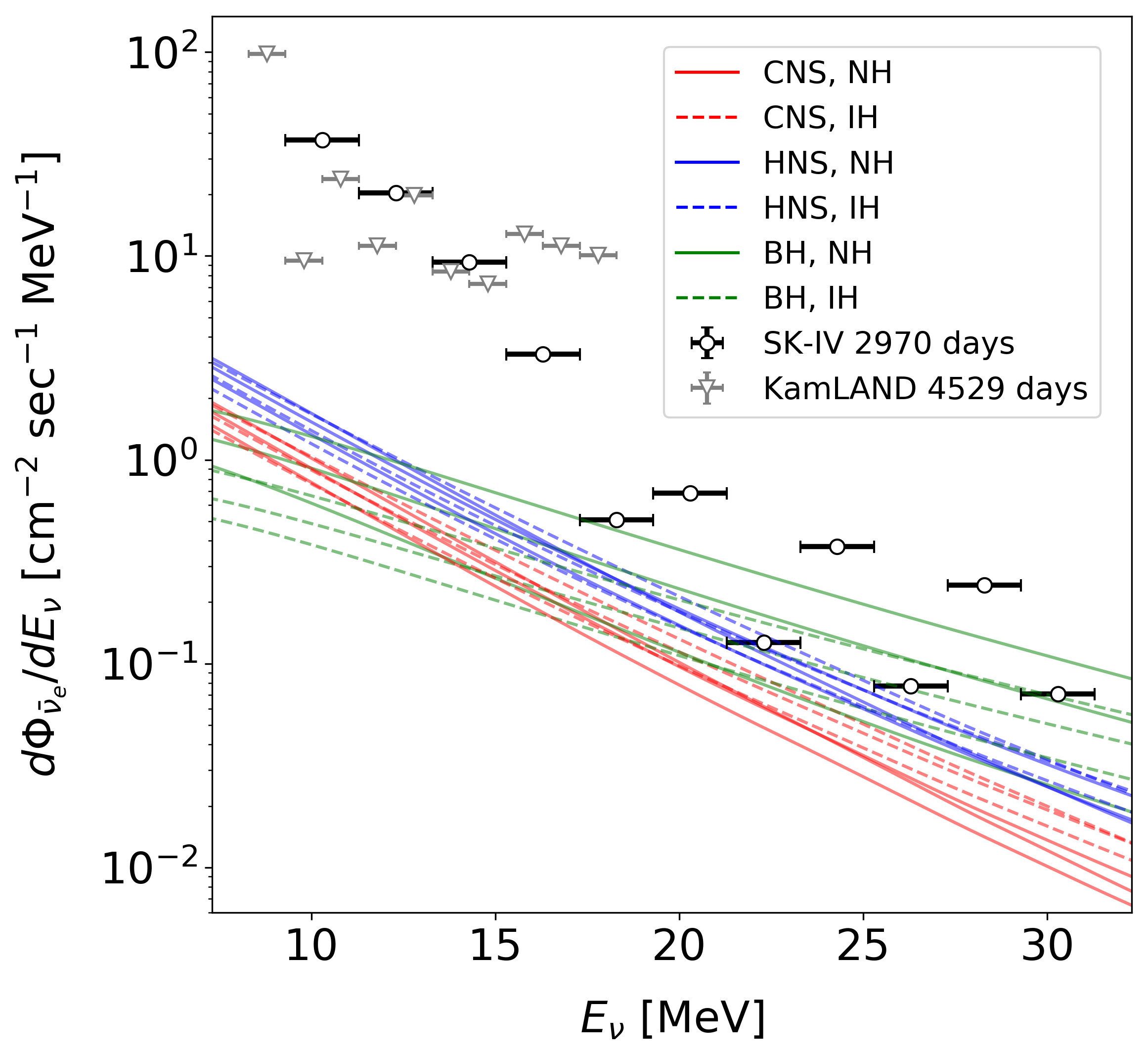}
  \end{center}
  \vspace{-10truept}
  \caption{Theoretical model predictions of the SRN $\bar\nu_e$ flux in the present study, in comparison with the 90\% confidence level (C.L.) upper limits on the extraterrestrial $\bar\nu_e$ flux from the latest experimental searches at SK~\citep[][]{2021arXiv210911174S} and KamLAND~\citep[][]{2022ApJ...925...14A}. The SRN $\bar\nu_e$ fluxes estimated under the assumption that neutrinos originate from a single component of the SN with a canonical-mass neutron star ($f_{\rm HNS}=0$ and $f_{\rm BH}=0$), the SN with a high-mass neutron star ($f_{\rm HNS}=1$ and $f_{\rm BH}=0$), and the failed SN ($f_{\rm BH}=1$) are shown with red, blue, and green lines, respectively. The solid and dashed lines correspond to the results for normal and inverted neutrino mass hierarchies (NH and IH), respectively. Here the fluxes with three different nuclear equations of state in Table~\ref{tab:snnmodel} are shown in the same style. Detailed descriptions about modeling are given in Section~\ref{sec:setup}.}
  \label{fig:fluxcomp}
  \end{figure}

Massive stars are likely to cause huge explosions with a gravitational collapse of their cores, as referred to as core-collapse supernovae (SNe). They are considered to be the major birthplace of neutrinos and therefore detecting neutrinos would promote understanding of the mechanism of SNe. Specifically, neutrinos could access physics at the most inner core, which is not possible via optical observations. So far, the detection of neutrinos from supernova bursts was only achieved for SN1987A~\citep{1987JETPL..45..589A,PhysRevLett.58.1494,PhysRevLett.58.1490}, because of a low supernova rate as well as detection difficulties of neutrinos. As a complementary strategy to the direct detection of neutrino bursts, measuring the accumulated flux from all distant SNe does not require us to wait for the rare explosion within a detectable range. This integrated flux is referred to as supernova relic neutrinos (SRNs), or the diffuse supernova neutrino background (DSNB) in some articles. The flux shape of SRNs depends on multiple factors, including not only the original neutrino spectrum from an explosion but also cosmic evolution, etc. Therefore, SRNs are believed to provide valuable information for further understanding of these physics. However, the observation has yet to be achieved, while a recent search at the Super-Kamiokande (SK) water Cherenkov detector has entered the range of theoretical predictions~\citep{2021arXiv210911174S}. Figure~\ref{fig:fluxcomp} shows the latest experimental upper limits from SK~\citep[][]{2021arXiv210911174S} and KamLAND~\citep[][]{2022ApJ...925...14A}, in comparison to predictions from our models in this paper. Various detectors such as water Cherenkov, liquid scintillators, xenon-based, and others in the next few decades are expected to achieve the SRN observation by improving sensitivity dramatically~\citep{2017JCAP...11..031P,2018JCAP...05..066M,2020PhRvD.102l3012D,2021PhRvD.103b3021S,2021arXiv220112920,PhysRevD.105.043008}. Accordingly, there is a higher need for the preparation to extract physical constraints when SRNs are actually detected. 

  %%%
  %%% 

In this study, we focus on the fate of stellar core collapse and its impact on the emitted neutrino spectrum. Total energy of neutrinos emitted from a core-collapse SN corresponds to the binding energy of a neutron star (NS) formed after the explosion \citep[][]{1987PhLB..196..267S,1989ApJ...340..426L}, which is determined by its mass and the nuclear equation of state (EOS). While the masses of NSs observed so far are typically $\sim1.35M_\odot$, high-mass NSs with $\gtrsim \! 1.6M_\odot$ were also discovered \citep[][]{2018MNRAS.478.1377A}. If a black hole (BH) is formed by core collapse, the spectrum of emitted neutrinos becomes different from the case of the SN with an NS and the total emission energy depends strongly on the EOS \citep[][]{2006PhRvL..97i1101S}. Therefore, the mass distribution of remnant NSs and the fraction of BH-forming collapses are responsible for the event rate of SRNs, and which factor is more largely responsible would depend on the nuclear EOS. Hence, measuring the SRN spectrum could provide implications about the core-collapse fate.\footnote{Note that uncertainties about many other factors affecting the SRN flux should be carefully considered to access the NS mass and BH formation fraction.}
 
This paper is structured as follows. We introduce recent theoretical studies and describe our models in Section~\ref{sec:setup} and give experimental knowledge and assumptions for the sensitivity study in Section~\ref{sec:bkg}. Then we present the results and discussion in Section~\ref{sec:result}. Finally, we conclude in Section~\ref{sec:concl}. 

%%---------------------------------------------------------------------------------
\section{Setup} \label{sec:setup}

\subsection{Formulation of the SRN flux} \label{sec:form}

The flux of SRNs is related to the evolution of galaxies that determines the cosmic rate of stellar core-collapse events \citep[][]{1996ApJ...460..303T}. In order to obtain the cosmic rate of stellar core-collapse events, we utilized the cosmological simulation results of the Illustris-1 run, which is the flagship of the Illustris simulation project \citep{2014MNRAS.444.1518V,2015A&C....13...12N}. The cosmic core-collapse rate at a redshift of $z$ is calculated as
  \begin{equation}
   R_{\rm CC}(z)=\dot\rho_\ast(z)\frac{\int^{M_{\rm max}}_{M_{\rm min}}\psi_{\rm IMF}(M)dM}{\int^{100M_\odot}_{0.1M_\odot}M\psi_{\rm IMF}(M)dM},
  \label{eq:ccrate}
  \end{equation}
where the cosmic star formation rate $\dot\rho_\ast(z)$ is adapted from the data of the Illustris-1 run which are publicly released for each snapshot at the selected redshift. Following the Illustris model \citep{2013MNRAS.436.3031V}, we choose the initial mass function (IMF) of progenitors $\psi_{\rm IMF}(M)$ proposed by \citet{2003PASP..115..763C}. Consequently, the maximum and minimum masses of progenitors that cause core collapse are set to $M_{\rm max}=100M_\odot$ and $M_{\rm min}=6M_\odot$, respectively. Note that the minimum mass for stars ending as a type II SN is assumed to be $6M_\odot$ in \citet{2013MNRAS.436.3031V} although larger values (8--$10M_\odot$) are adopted for $M_{\rm min}$ in the previous SRN studies \citep{2015ApJ...804...75N,2018MNRAS.475.1363H,2021ApJ...909..169K}. This leads to the cosmic core-collapse rate at $z=0$ of $R_{\rm CC}(0)=3.9\times 10^{-4}\,{\rm Mpc}^{-3}\,{\rm yr}^{-1}$, which is significantly higher than the typical value of the core-collapse SN rate obtained from observations \citep[e.g.,][]{2014ARA&A..52..415M}. Nevertheless, the cosmic core-collapse rate of around $z = 1$ is comparable to that of \citet{2014ApJ...790..115M}, which is often adopted in previous SRN studies. A recent study using the Gaia DR2 database suggests that the upper limit on white dwarf formation is $\sim6M_\odot$ \citep[][]{2021ApJ...912..165R}.\footnote{According to the latest study by \citet{2022ApJ...926L..24M}, a candidate white dwarf with a precursor mass of $8.5M_\odot$ is found from the Gaia EDR3 catalog.} Based on these settings, we obtain $R_{\rm CC}(z)\simeq\dot\rho_\ast(z)/(57.5M_\odot)$, $\dot\rho_\ast(z)/(84.6M_\odot)$, and $\dot\rho_\ast(z)/(115M_\odot)$ for $M_{\rm min}=6M_\odot$, $8M_\odot$, and $10M_\odot$, respectively. On the contrary, according to \citet{2021PhRvD.103d3003H}, mergers of binary stars whose individual masses are originally below the core-collapse threshold can produce SN progenitors and enhance the SRN flux. In addition, the redshift evolution of the IMF could be taken into account for the estimation of the SRN event rate \citep[][]{2021arXiv211102624A}.

We classify the fate of stellar collapse into four types: the SN with a low-mass NS, the SN with a high-mass NS, the faint SN, and the failed SN. The low- and high-mass NSs are divided by a baryon mass of $M_{\rm NS, b}=1.6M_\odot$.\footnote{According to the approximation by \citet{1989ApJ...340..426L}, neutron stars with $M_{\rm NS, b}=1.6M_\odot$ are expected to have a gravitational mass of $M_{\rm NS, g} \approx 1.43M_\odot$. Note that the relation between the baryon mass and gravitational mass of neutron stars depends on the EOS while the accuracy of this approximation would be at most $\pm$2\%.} Faint SNe are successful explosions accompanied by a substantial fallback that leads to a BH formation at later times. Since they are considered to be rare, we will not take into account their contribution in the following study of this paper. In failed SNe, a BH is formed by continuous mass accretion without explosion. The fate of stellar core collapse strongly affects the neutrino emission. In the previous SRN studies \citep{2015ApJ...804...75N,2018MNRAS.475.1363H,2021ApJ...909..169K}, the fate is assumed to be determined by the progenitor model, and the IMF-weighted neutrino spectrum is used to calculate the SRN flux. Nevertheless, this approach may be inconvenient because the evolutionary scenario for progenitors and explosion mechanism of core-collapse SNe are still uncertain. In other words, the stellar core-collapse fate depends not only on the progenitor mass and metallicity but also on the evolutionary model.

In the present study, we consider three cases for the fate of stellar core collapse: the SN with a canonical-mass NS, the SN with a high-mass NS, and the failed SN forming a BH.\footnote{After the first submission of this paper, \citet{2022arXiv220605299E} showed the prediction of SRN event rate focusing on the NS mass.} Consequently, we set their fractions directly as parameters and define the fate-weighted average neutrino number spectrum as
  \begin{eqnarray}
   \left\langle \frac{dN(E^\prime_\nu)}{dE^\prime_\nu} \right\rangle 
    = f_{\rm BH}\frac{dN_{\rm BH}(E^\prime_\nu)}{dE^\prime_\nu} 
    + (1-f_{\rm BH}) \nonumber \\ \times \left[ f_{\rm HNS}\frac{dN_{\rm HNS}(E^\prime_\nu)}{dE^\prime_\nu} + (1-f_{\rm HNS})\frac{dN_{\rm CNS}(E^\prime_\nu)}{dE^\prime_\nu} \right],
  \label{eq:avespct}
  \end{eqnarray}
where $f_{\rm BH}$ is the fraction of failed SNe to the total number of stellar core-collapse events and $f_{\rm HNS}$ is the fraction of SNe with a high-mass NS to the total number of SNe with an NS. Furthermore, we employ generic models of the neutrino spectrum $dN_{\rm CNS}(E^\prime_\nu)/dE^\prime_\nu$, $dN_{\rm HNS}(E^\prime_\nu)/dE^\prime_\nu$, and $dN_{\rm BH}(E^\prime_\nu)/dE^\prime_\nu$ for the SN with a canonical-mass NS, the SN with a high-mass NS, and the failed SNe with a BH, respectively. We then describe the SRN flux as 
  \begin{eqnarray}
   \frac{d\Phi(E_\nu)}{dE_\nu}=c\int^{z_{\rm max}}_{0} R_{\rm CC}(z) \left\langle \frac{dN(E^\prime_\nu)}{dE^\prime_\nu} \right\rangle \nonumber \\ \times  \frac{dz}{H_0\sqrt{\Omega_{\rm m}(1+z)^3+\Omega_\Lambda}},
  \label{eq:flux}
  \end{eqnarray}
where $c$ is the speed of light and the cosmological constants are $\Omega_{\rm m}= 0.2726$, $\Omega_\Lambda= 0.7274$, and $H_0= 70.4$~km~s$^{-1}$~Mpc$^{-1}$ \citep{2014MNRAS.444.1518V}. Here, the neutrino energy at a detector $E_\nu$ is related to that at a source $E^\prime_\nu$ as $E^\prime_\nu=(1+z)E_\nu$ when the source is located at a redshift $z$.

The generic models of the neutrino spectrum for each component are adopted from the results of numerical simulations. As for the SN with an NS, we refer to \citet{2022ApJ...925...98N} and employ the models with NS baryonic masses of $M_{\rm NS, b}=1.47M_\odot$ and $1.86M_\odot$ for the representatives of a canonical-mass NS and a high-mass NS, respectively. Furthermore, in order to obtain the time-integrated neutrino spectra, we prepare the neutrino light curves from an onset of core collapse to cooling of the proto-NS by following the method adopted in the Supernova Neutrino Database \citep{2013ApJS..205....2N}. The shock revival time used in this procedure is taken from Table 1 of \citet{2022ApJ...925...98N} while it is denoted as $t_{\rm init}$. The progenitors of our SN models are taken from \citet{1995ApJS..101..181W}; the models with $M_{\rm NS, b}=1.47M_\odot$ and $1.86M_\odot$ have progenitors of the solar metallicity with zero-age main-sequence masses of $15M_\odot$ and $40M_\odot$, respectively. On the other hand, for the failed SN, we utilize the models in \citet{2021PASJ...73..639N}, where the core collapse and neutrino emission are studied for the same progenitor model (an initial mass of $30M_\odot$ and metallicity of $Z=0.004$) involving a BH formation as in \citet{2013ApJS..205....2N}.

The neutrino signal from stellar core collapse depends on the nuclear EOS. In the case of the SN with an NS, the total energy of emitted neutrinos is determined by the gravitational binding energy of the remnant NS and the EOS affects the average energy of neutrinos \citep{2018PhRvC..97c5804N} and the cooling timescale of the proto-NS \citep{2019ApJ...878...25N,2020ApJ...891..156N}. In particular, although the Supernova Neutrino Database \citep{2013ApJS..205....2N} provides the neutrino spectra up to 20~sec, the neutrino emission continues for a longer period \citep{2019ApJ...881..139S} and the EOS dependence gets stronger for the later phase \citep[][]{2022ApJ...925...98N}. In the case of the failed SN, the time-integrated neutrino spectrum depends on the maximum mass of the proto-NS because neutrinos are emitted until a BH formation \citep{2006PhRvL..97i1101S}. In the previous studies \citep{2009PhRvL.102w1101L,2015ApJ...804...75N,2020MPLA...3530011M}, the EOS dependence is taken into account only for the component of the failed SN. In the present study, we also consider the EOS dependence for the SN with an NS. For this purpose, we employ three EOS models: Togashi \citep{2017NuPhA.961...78T}, LS220 \citep{1991NuPhA.535..331L}, and Shen \citep{2011ApJS..197...20S}. Then, the neutrino spectra obtained with use of different EOS models are prepared for each component. %Table~\ref{tab:snnmodel} shows the entries of neutrino signal adopted in the present study.

We show the average and total energies of our time-integrated neutrino signal in Table~\ref{tab:snnmodel}. First, we compare the models with different fates. The average neutrino energy of the SN with a high-mass NS is almost the same as that of the SN with a canonical-mass NS. In contrast, the models with a high-mass NS give a larger total energy of neutrinos than those with a canonical-mass NS because the remnant NS has a larger binding energy available for the neutrino emission. As for the failed SN, the gravitational potential energy released from core collapse is not fully converted to the neutrino radiation since a BH is formed. Thus the total neutrino energy of the failed SN is comparable to that of the SN with an NS. Nevertheless, the failed SN has a higher average energy of neutrinos than the SN with an NS because heating due to matter accretion continues until a BH formation. Next, we consider the EOS dependence. Since the total energy of emitted neutrinos corresponds to a binding energy of the remnant NS, EOS models with a smaller NS radius have a larger total energy for the SN with an NS. Among our models, the Togashi and Shen EOSs have the smallest and largest NS radii, respectively. In the case of the failed SN, the amount of emitted neutrinos is larger for EOS models with a higher maximum mass because the time taken to form a BH and accordingly duration of the neutrino emission are longer. Note that the maximum mass of proto-NSs is different from that of cold NSs and depends on the entropy as shown in Figure~13 of \citet{2020ApJ...894....4D}.

  %%%
  \begin{deluxetable*}{llrrrrrrrrr}
  \tablecaption{Properties of the compact remnant and time-integrated neutrino signal for each fate component\label{tab:snnmodel}.}
  \tablewidth{0pt}
  \tablehead{
  \colhead{} & \colhead{} & \colhead{$M_{\rm NS,g}$} & \colhead{$R_{\rm NS}$} & \colhead{$t_{\rm BH}$} & \colhead{$\langle E_{\nu_e} \rangle$} & \colhead{$\langle E_{\bar \nu_e} \rangle$} & \colhead{$\langle E_{\nu_x} \rangle$} & \colhead{$E_{\nu_e, {\rm tot}}$} & \colhead{$E_{\bar \nu_e, {\rm tot}}$} & \colhead{$E_{\nu_x, {\rm tot}}$} \\
  \colhead{EOS} & \colhead{Remnant} & \colhead{($M_\odot$)} & \colhead{(km)} & \colhead{(sec)} & \multicolumn3c{(MeV)} & \multicolumn3c{($10^{52}$ erg)}
  }
  \startdata
  Togashi & canonical-mass NS & 1.32 & 11.5 & --- & 9.2 & 10.9 & 11.8 & 4.47 & 4.07 & 4.37 \\
   & high-mass NS & 1.63 & 11.5 & --- & 9.5 & 11.2 & 11.9 & 7.26 & 6.93 & 7.17 \\
   & BH (Failed SN) & --- & --- & 0.533 & 16.1 & 20.4 & 23.4 & 6.85 & 5.33 & 2.89 \\ \hline
  LS220 & canonical-mass NS & 1.34 & 12.7 & --- & 9.1 & 10.7 & 11.3 & 4.25 & 3.84 & 3.94 \\
   & high-mass NS & 1.65 & 12.4 & --- & 9.8 & 11.2 & 11.2 & 7.29 & 6.88 & 6.36 \\
   & BH (Failed SN) & --- & --- & 0.342 & 12.5 & 16.4 & 22.3 & 4.03 & 2.87 & 2.11 \\ \hline
  Shen & canonical-mass NS & 1.35 & 14.3 & --- & 9.0 & 10.6 & 11.3 & 3.65 & 3.22 & 3.35 \\
   & high-mass NS & 1.67 & 14.1 & --- & 9.6 & 11.2 & 11.2 & 6.22 & 5.88 & 5.40 \\
   & BH (Failed SN) & --- & --- & 0.842 & 17.5 & 21.7 & 23.4 & 9.49 & 8.10 & 4.00 \\
  \enddata
  \tablecomments{For the successful SN models, $M_{\rm NS,g}$ and $R_{\rm NS}$ are the gravitational mass and radius of NSs, respectively. For the failed SN models, $t_{\rm BH}$ is the time to a BH formation measured from the core bounce. $\langle E_{\nu_i} \rangle$ and $E_{\nu_i, {\rm tot}}$ are the average and total energies of the time-integrated neutrino signal for $\nu_i$, where $\nu_x$ represents the average of $\nu_\mu$, $\bar \nu_\mu$, $\nu_\tau$, and $\bar \nu_\tau$.}
  \end{deluxetable*}
  %%%

\subsection{Distribution of compact remnants} \label{sec:distcm}

In our study, the distribution of compact remnants formed by stellar collapse is encapsulated in two parameters, $f_{\rm BH}$ and $f_{\rm HNS}$. Observationally, a bimodal mass distribution is inferred for the NS in the binary system \citep{2011MNRAS.414.1427V,2012ApJ...757...55O,2013ApJ...778...66K,2016arXiv160501665A,2018MNRAS.478.1377A}. The low-mass component of the distribution contains a majority of the sample and has a narrow peak at $\sim$1.35$M_\odot$. Canonical-mass NS models in our calculation are chosen to be consistent with this component (Table~\ref{tab:snnmodel}). On the other hand, the high-mass component, which mainly consists of the NS with a white dwarf companion, has a broad peak around 1.5--$1.8M_\odot$. The bimodal mass distribution is interpreted as a result of difference in the birth mass and/or additional accretion history. \citet{2016arXiv160501665A} claimed that a certain fraction of the observed high-mass NSs should have been born with a mass larger than $1.6M_\odot$ because the accretion efficiency from the binary companion was estimated to be low. In contrast, a stellar evolution model by \citet{2020ApJ...896...56W} produced much fewer high-mass NSs than claimed by \citet{2016arXiv160501665A}. Incidentally, \citet{2020ApJ...896...56W} shows that the birth mass function of NSs is insensitive to the mass-loss rate. In our calculation, $f_{\rm HNS}$ corresponds to the fraction of the high-mass NS at birth, which is still not well constrained. 

Failed SNe associating a BH formation would be observed as a disappearance of luminous stars without SNe in optical wavelengths \citep{2008ApJ...684.1336K}. Recently, \citet{2021arXiv210403318N} compiled 11 yr of the monitoring data for luminous stars in nearby galaxies and found the failed SN fraction to be 4--39\% at a 90\% C.L. provided that one failed SN was detected. This range is consistent with \citet{2020ApJ...896...56W}, whose model predicts a fraction of 9--32\%. In this prediction, the fraction of BH-forming collapses is sensitive to the mass-loss rate. Because the mass-loss rate depends on the stellar metallicity, the failed SN fraction would be a function of the redshift; $f_{\rm BH}$ is considered to be larger for higher redshifts in general \citep{2013PhRvD..88h3012N,2015ApJ...804...75N,2015PhLB..751..413Y}. Nevertheless, we assume that the fraction of failed SNe does not depend on redshift for simplicity.

\subsection{Neutrino oscillation and detection} \label{sec:oscdtc}

In water Cherenkov detectors, the main detection channel of SRNs is inverse beta decay (IBD) of electron antineutrinos:
  \begin{equation}
   \bar \nu_e + p \to e^+ + n .
  \label{eq:ibd}
  \end{equation}
Therefore, we focus on the flux of $\bar\nu_e$ throughout the article. For the models of neutrino spectrum described above (Table~\ref{tab:snnmodel}), the neutrino flavor oscillations are not taken into account. The terrestrial flux of $\bar\nu_e$ is written as
  \begin{equation}
   \frac{d\Phi^{\rm det}_{\bar\nu_e}(E_\nu)}{dE_\nu}=\bar P \frac{d\Phi_{\bar\nu_e}(E_\nu)}{dE_\nu} +(1-\bar P)\frac{d\Phi_{\nu_x}(E_\nu)}{dE_\nu},
  \label{eq:fluxobs}
  \end{equation}
where $d\Phi_{\bar\nu_e}(E_\nu)/dE_\nu$ and $d\Phi_{\nu_x}(E_\nu)/dE_\nu$ are fluxes of $\bar\nu_e$ and $\nu_x$ at a core collapse, respectively. Note that we denote $\nu_\mu$, $\bar \nu_\mu$, $\nu_\tau$, and $\bar \nu_\tau$ as $\nu_x$ collectively because the difference in their emission processes is subtle. The effect of flavor conversions is encapsulated in a survival probability of $\bar\nu_e$, $\bar P$. Following \citet{2015ApJ...804...75N}, we consider the flavor conversions due to the matter-induced effect \citep[the so-called the MSW effect;][]{1978PhRvD..17.2369W,1985YaFiz..42.1441M} that takes place in the stellar envelope. The flavor mixing depends on the neutrino mass hierarchy \citep{2000PhRvD..62c3007D} as $\bar P \approx 0.7$ for the normal mass hierarchy (NH) and $\bar P \approx 0$ for the inverted mass hierarchy (IH). Also the self-induced (collective) effect can exchange $\bar\nu_e$ and $\nu_x$ \citep[e.g.,][]{2010ARNPS..60..569D}. Unfortunately, the self-induced conversion is complicated and still unknown because it is hard to solve the flavor evolution involving the self-interaction terms in general. Thereby, the oscillation probabilities depend on the time and energy. Nevertheless, according to \citet{2012JCAP...07..012L}, the self-induced effect on the SRN spectrum is minor. In the present study, we do not consider the collective oscillation while further investigations are important.

The SRN event rate is calculated as
  \begin{equation}
   \frac{dN_{\rm event}(E_{e^+})}{dE_{e^+}}=N_p\cdot\sigma_{\rm IBD}(E_\nu) \cdot\frac{d\Phi^{\rm det}_{\bar\nu_e}(E_\nu)}{dE_\nu},
  \label{eq:evrate}
  \end{equation}
where $\sigma_{\rm IBD}(E_\nu)$ is the IBD cross section taken from \citet{2003PhLB..564...42S}. For simplicity, we set the positron energy to be $E_{e^+} = E_\nu-\Delta c^2$ with a neutron-proton mass difference, $\Delta$, while the recoil effects that are provided in \citet{2003PhLB..564...42S} also reduce the positron energy. $N_p$ represents the number of free protons contained in the detector.

%%---------------------------------------------------------------------------------
\section{Detectors and Background Sources} \label{sec:bkg}

In this study, we focus on water-based Cherenkov detectors. This type of detectors includes SK, its upgrade with a gadolinium loading, SK-Gd, and the future large volume detector, Hyper-Kamiokande (HK). We consider the SRN $\bar\nu_e$ search range and background sources at SK-Gd and HK based on the latest SK analysis with 2970 days of data presented in \citet{2021arXiv210911174S}, as they share many in common. The detector volume of SK-Gd is set to the same as SK ($N_p=1.5\times 10^{33}$) while that of HK is set to 8.4 times larger ($N_p=12.6\times 10^{33}$) as assuming their fiducial volumes. As a detector operation time, we consider 10 yr for SK-Gd, and 3, 5, and 10 yr for HK. The backgrounds are scaled by the assumed livetime and detector volume from \citet{2021arXiv210911174S}.

The detection channel is IBD of electron antineutrinos as mentioned above. At SK, positrons are detected via Cherenkov light emitted from their travel in the detector. Neutrons cannot be detected directly as they are electromagnetically neutral, but SK made use of a 2.2~MeV $\gamma$ ray emitted from their capture on hydrogens after thermalization. Since the signal size of electromagnetic cascades from this 2.2~MeV $\gamma$ ray is small and then difficult to distinguish from a huge amount of low energy backgrounds, a machine-learning technique was employed in the analysis. The signal efficiency with this method was 20$-$30\%, depending on energy region, to reduce low backgrounds to a manageable size.\footnote{We scale from the 30\% neutron tagging at SK, as the efficiency around this was achieved for the chosen criteria for the energy region of current interest in \citet{2021arXiv210911174S}.} At SK-Gd, because $\gamma$ rays emitted from neutron captures on gadolinium have a higher total energy ($\sim$8~MeV) enough to be detected unambiguously, the neutron tagging efficiency is expected to be dramatically improved, depending on the gadolinium loading rate.\footnote{The capture rate is $\sim$50\% with 0.01\% gadolinium in water and $\sim$90\% with 0.1\% according to \citet{2021arXiv220112920}.} In this paper, we assume a 70\% neutron tagging efficiency for SK-Gd, while a higher efficiency may be achieved in the experiment. As for HK, since there are still many unclear factors, we assume the same tagging efficinecy as that for SK.\footnote{The photosensors for HK have a better performance than those for SK, but the neutron tagging performance depends also on the detector size, the water purity, the coverage of sensors, etc.} We assume the energy resolutions at both of SK-Gd and HK are the same as that at SK. 

The analysis energy window for the SRN search was set to $9.3 < E_\nu < 31.3$~MeV in \citet{2021arXiv210911174S}. In this paper, we show the results from two different energy ranges: $13.3 < E_\nu < 31.3$~MeV and $17.3 < E_\nu < 31.3$~MeV. The background source has its characteristic energy region in the search window; reactor neutrinos, solar neutrinos, and cosmic ray muon spallation events dominate in the lower side ($E_\nu \lesssim 17.3$~MeV), while the atmospheric neutrino background becomes an almost only one contributor at higher energies ($E_\nu \gtrsim 17.3$~MeV). 

There are two types of spallation backgrounds. One is isotopes, such as $^9$Li, that make almost identical signitures as IBD via $\beta+n$ decay. The other is isotopes whose decay products do not include neutrons. Most of these can be easily rejected by the neutron tagging technique, but some make an accidental pair with a 2.2~MeV $\gamma$-ray-like signal and then mimic IBD events. These accidental backgrounds are expected to be reduced dramatically at SK-Gd, thanks to much better visuality of $\sim$8~MeV $\gamma$ rays. We therefore assume a 10 times smaller amount of accidental backgrounds at SK-Gd. In contrast, cosmic ray muon spallation may increase at HK as overburden at the planned construction mountain is smaller than that at the SK site. Nevertheless, since there are still a lot of unclear factors, we assume the same amount of spallation at HK. In the later study, note that the background assumption for the results with $13.3 < E_\nu < 31.3$~MeV would be underestimated for this reason. The results with $17.3 < E_\nu < 31.3$~MeV are not affected as this region is almost spallation-free. 

In the SRN search at SK, atmospheric neutrino backgrounds are classified into two categories as neutral-current quasielastic (NCQE) interactions and the others (non-NCQE). NCQE interactions involve nuclear $\gamma$-ray emission via primary neutrino and following neutron reactions \citep[][]{PhysRevD.100.112009}, which mimic the positron signal from IBD. Charged-current interactions of atmospheric electron antineutrinos give the exactly same final state as IBD. Charged-current muon neutrino interactions would become backgrounds when the final state muons have less energy than the Cherenkov threshold but electrons from their decays are visible in the detector. Some of neutral-current interactions involve a low energy pion, which is also invisible while an electron from its decay is visible, and then look similar to signal. We assume the same fractions of these atmospheric backgrounds at SK-Gd and HK as SK. 

We estimate the expected sensitivity to the SRN $\bar\nu_e$ flux by using pseudo experiments following the method in \citet{2021arXiv210911174S}; we vary background events by their statistical and systematic uncertainties then obtain the upper bound on the number of events at a certain C.L., $N_{\rm lim}$. The expected upper bound on the $\bar\nu_e$ flux is then calculated as
  \begin{equation}
   \Phi_{\rm lim} = \frac{N_{\rm lim}}{T \cdot N_p \cdot \bar{\sigma}_{\rm IBD} \cdot \epsilon_{\rm sig}}, 
  \label{eq:snstv}
  \end{equation}
where $T$ is the livetime of the detector operation, $\bar{\sigma}_{\rm IBD}$ is the IBD cross section at a mean energy in the region, and $\epsilon_{\rm sig}$ is the signal efficiency. We use the IBD cross section at 24.3 (22.3)~MeV for the analysis with an energy range of $17.3 \ (13.3) < E_\nu < 31.3$~MeV. This treatment may cause a change in the result for a nonlinear shape of the differential cross section, but the impact on the final results is limited. Here we assume multiple assumptions on systematic uncertainties. In comparison to a 60\% uncertainty at SK, we use a $\times0.7$ size of uncertainty on NCQE for SK-Gd 10 yr. On the other hand, we assume 30\%, 20\%, and 10\% for HK 3 yr, HK 5 yr, and HK 10 yr, respectively.\footnote{This uncertainty in the SK analysis was energy dependent in \citet{2021arXiv210911174S}, but we scale from 60\% as an uncertainty size in the current energy range of interest.} These are based on the expected efforts in accelerator neutrino and nuclear experiments as described in \citet{PhysRevD.100.112009}. In \citet{2021arXiv210911174S}, the non-NCQE background was estimated by using the sideband region, whose uncertainty is then mostly statistical. We assume a $\times0.7$ size of uncertainty compared to the SK one on this for SK-Gd 10 yr, while use 10\% for HK 3 yr, 8\% for HK 5 yr, and 5\% for HK 10 yr. We take the same size of systematic uncertainties for the other backgrounds from \citet{2021arXiv210911174S}. The uncertainties for these other backgrounds may be improved in the future, but the impact is subdominant in the current study. Each condition for different detectors is summarized in Table~\ref{tab:detecsummary}. 

  %%%
  \begin{deluxetable*}{cccccc}
  \tablecaption{Experimental conditions assumed for the present sensitivity study\label{tab:detecsummary}.}
  \tablewidth{0pt}
  \tablehead{
  \colhead{} & SK-IV 2970 days & SK-Gd 10 yr & HK 3 yr & HK 5 yr & HK 10 yr
  }
  \startdata
    Neutron tagging efficiency & 30\% & 70\% & 30\% & 30\% & 30\% \\ 
    Accidental background amount & $-$ & $\times$0.1 & same & same & same \\ 
    NCQE uncertainty size & 60\% & 42\% & 30\% & 20\% & 10\% \\ 
    Non-NCQE uncertainty size & 19\% & 14\% & 10\% & 8\% & 5\% \\ 
  \enddata
  \tablecomments{The SK-IV case is shown as a reference. The efficiencies and uncertainties are actually energy dependent in the SK-IV analysis, while we assume representative values and extrapolate to each case from these.}
  \end{deluxetable*}
  %%%

%%---------------------------------------------------------------------------------
\section{Results} \label{sec:result}

\subsection{Flux and event rate for each fate} \label{sec:cmpnt}

Here we discuss the SRN fluxes evaluated under the assumption that SRNs originate from a single component listed in Table~\ref{tab:snnmodel}. In relation to Equation~(\ref{eq:avespct}), the models of the SN with a canonical-mass NS, the SN with a high-mass NS, and the failed SN with a BH correspond to $(f_{\rm HNS},f_{\rm BH})=(0,0)$, $(f_{\rm HNS},f_{\rm BH})=(1,0)$, and $f_{\rm BH}=1$, respectively. The $\bar\nu_e$ fluxes of SRNs obtained in this way are shown as a function of neutrino energy in Figure~\ref{fig:fluxcomp}, where the 90\% C.L. observed upper limits on the extraterrestrial $\bar\nu_e$ flux from the latest experimental searches at SK \citep[][]{2021arXiv210911174S} and KamLAND \citep[][]{2022ApJ...925...14A} are also shown for comparison. Whereas some models of the failed SN are disfavored by the observed limits, a feasible flux should be obtained by mixing all three components as in Equation~(\ref{eq:avespct}).

\begin{figure*}[htbp]
    \centering
    \begin{minipage}{0.35\hsize}
        \begin{center}
            \includegraphics[clip,width=7.0cm]{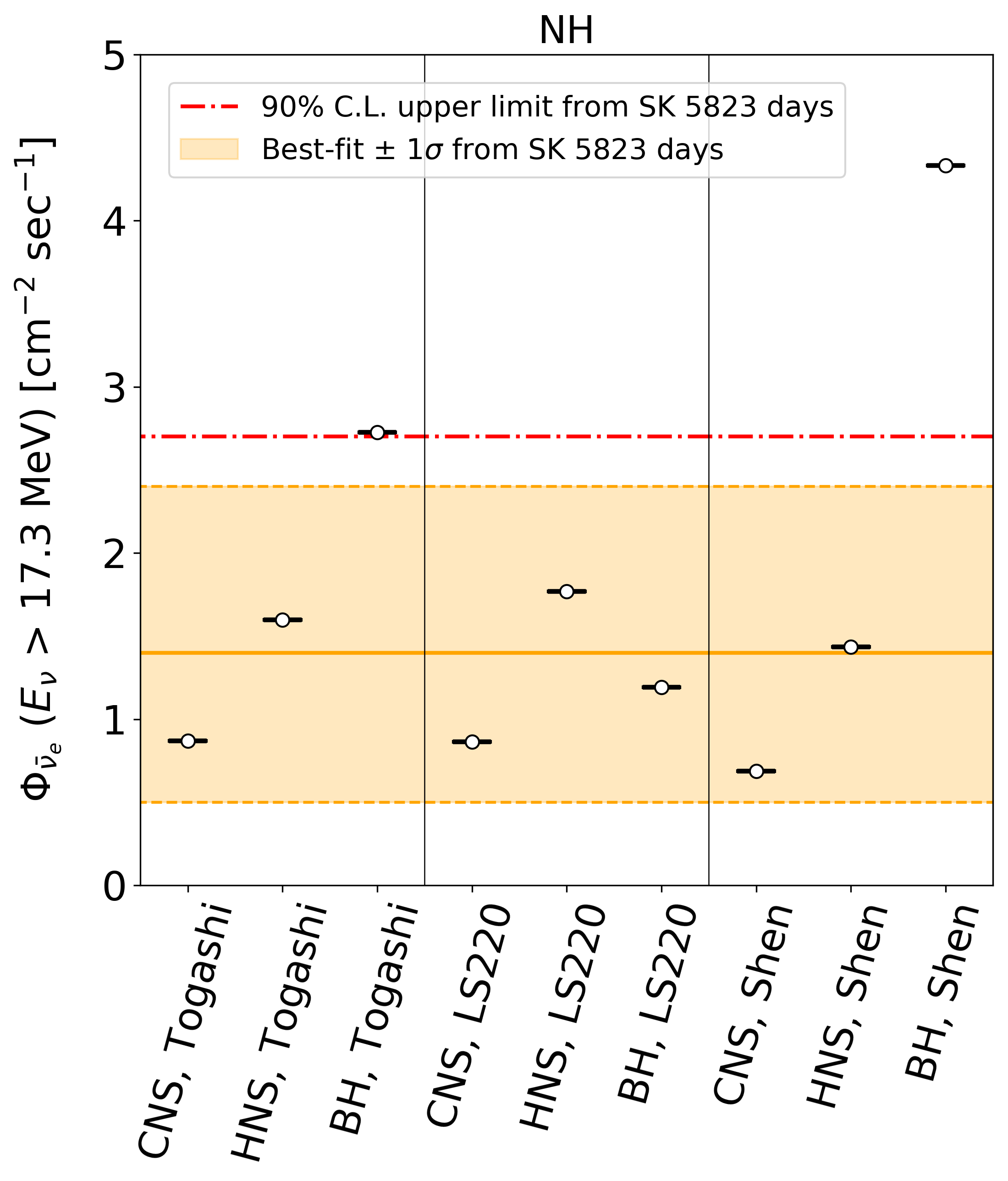}
        \end{center}
    \end{minipage}
    \hspace{+50truept}
    \begin{minipage}{0.35\hsize}
        \begin{center}
            \includegraphics[clip,width=7.0cm]{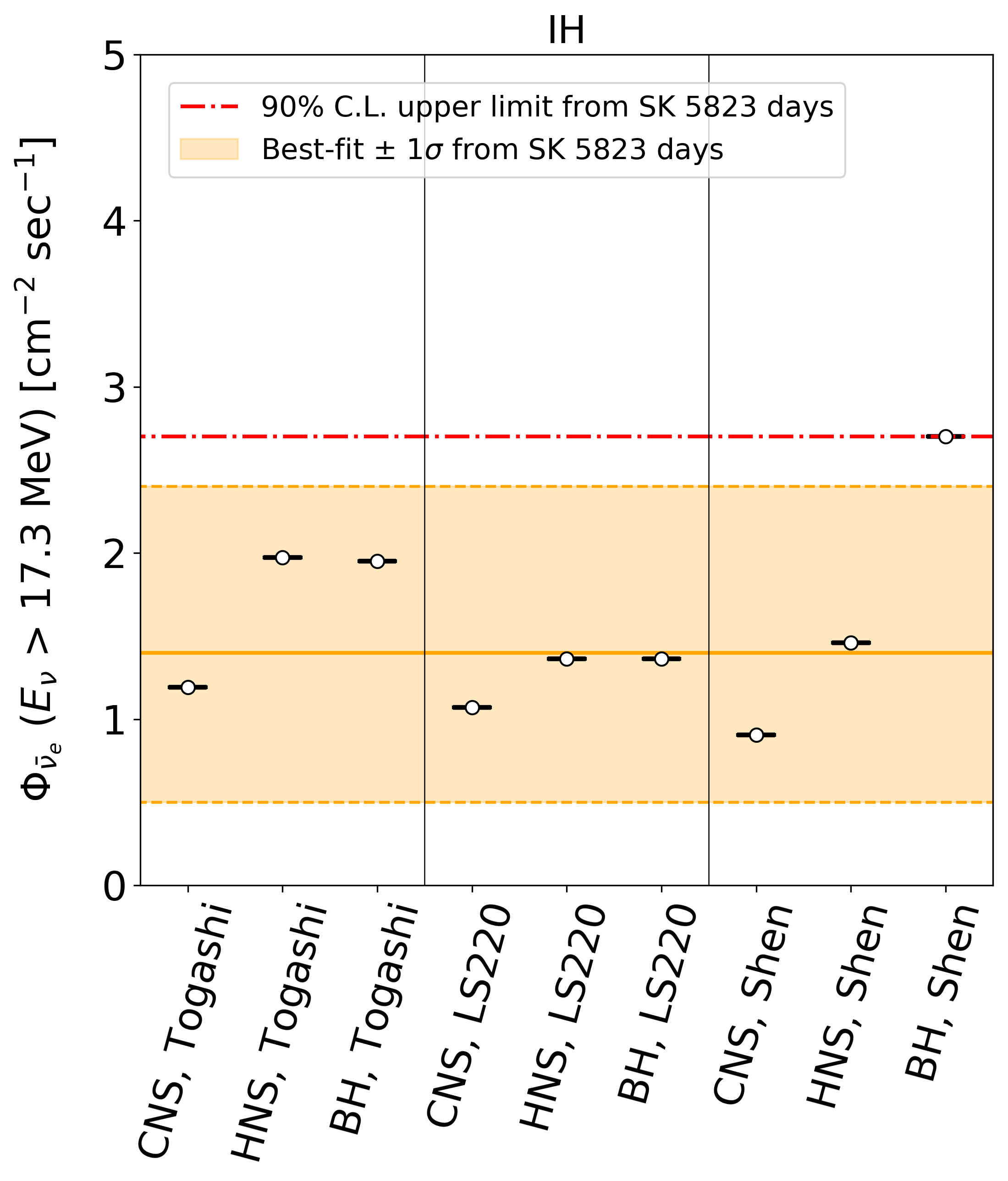}
        \end{center}
    \end{minipage}
    \vspace{+10truept}
    \caption{The integrated SRN $\bar\nu_e$ fluxes with $E_\nu > 17.3~{\rm MeV}$ estimated under the assumption that SRNs originate from a single component listed in Table~\ref{tab:snnmodel}. The left and right panels show the results for NH and IH, respectively. The orange solid lines and shaded regions represent the best-fit values and their $\pm 1\sigma$ uncertainties and the red dashed-dotted lines the 90\% observed upper limits from the latest SK analysis \citep[][]{2021arXiv210911174S} associated with the spectral models in \citet{2015ApJ...804...75N}. Note that the SRN flux for each component may change for various assumptions such as $M_{\rm min}$ and BH formation model as discussed in Section~\ref{sec:discuss}.}
    \label{fig:integcomp}
  \end{figure*}

In Figure~\ref{fig:integcomp}, the integrated SRN $\bar\nu_e$ fluxes with $E_\nu > 17.3~{\rm MeV}$ for the single component models are shown together with the best-fit results and 90\% C.L. upper limits in \citet{2021arXiv210911174S}.\footnote{In \citet{2021arXiv210911174S}, the best-fit values with $\pm 1\sigma$ and the corresponding upper limits were evaluated assuming normalized spectra taken from several theoretical models. While we show in Figure~\ref{fig:integcomp} the best-fit result associated with the models of \citet{2015ApJ...804...75N}, $1.4^{+1.0}_{-0.9}\,{\rm cm}^{-2}\,{\rm sec}^{-1}$ (same for both of the NH and IH cases), this value is insensitive to the normalized spectrum provided that realistic models are adopted.} For the SN models with an NS, the integrated flux varies by the EOS and remnant and its dependence is similar to that of the total energy of emitted neutrinos in Table~\ref{tab:snnmodel}. On the other hand, the failed SN models have a larger flux in $E_\nu > 17.3~{\rm MeV}$ for their total emission energy because their average neutrino energy is higher than that of the SN models with an NS. Note that the fluxes of the present theoretical models are larger than those of \citet{2015ApJ...804...75N} \citep[see also Figure~29 of][]{2021arXiv210911174S} because the conversion factor from $\dot\rho_\ast(z)$ to $R_{\rm CC}(z)$ is larger. 

  %%%
  %%%

\begin{figure*}[htbp]
  \gridline{\fig{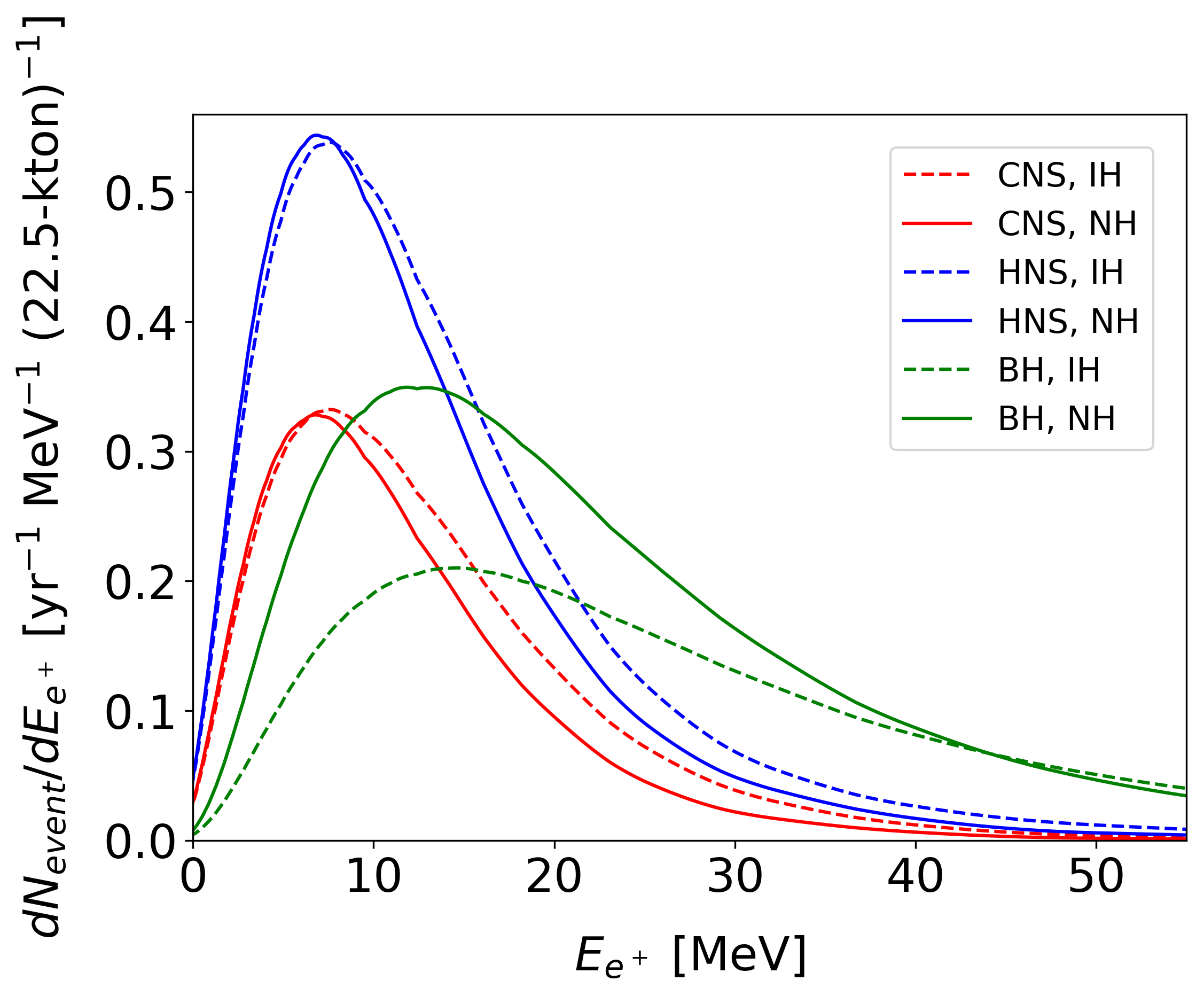}{0.33\textwidth}{(a) Togashi EOS}
            \fig{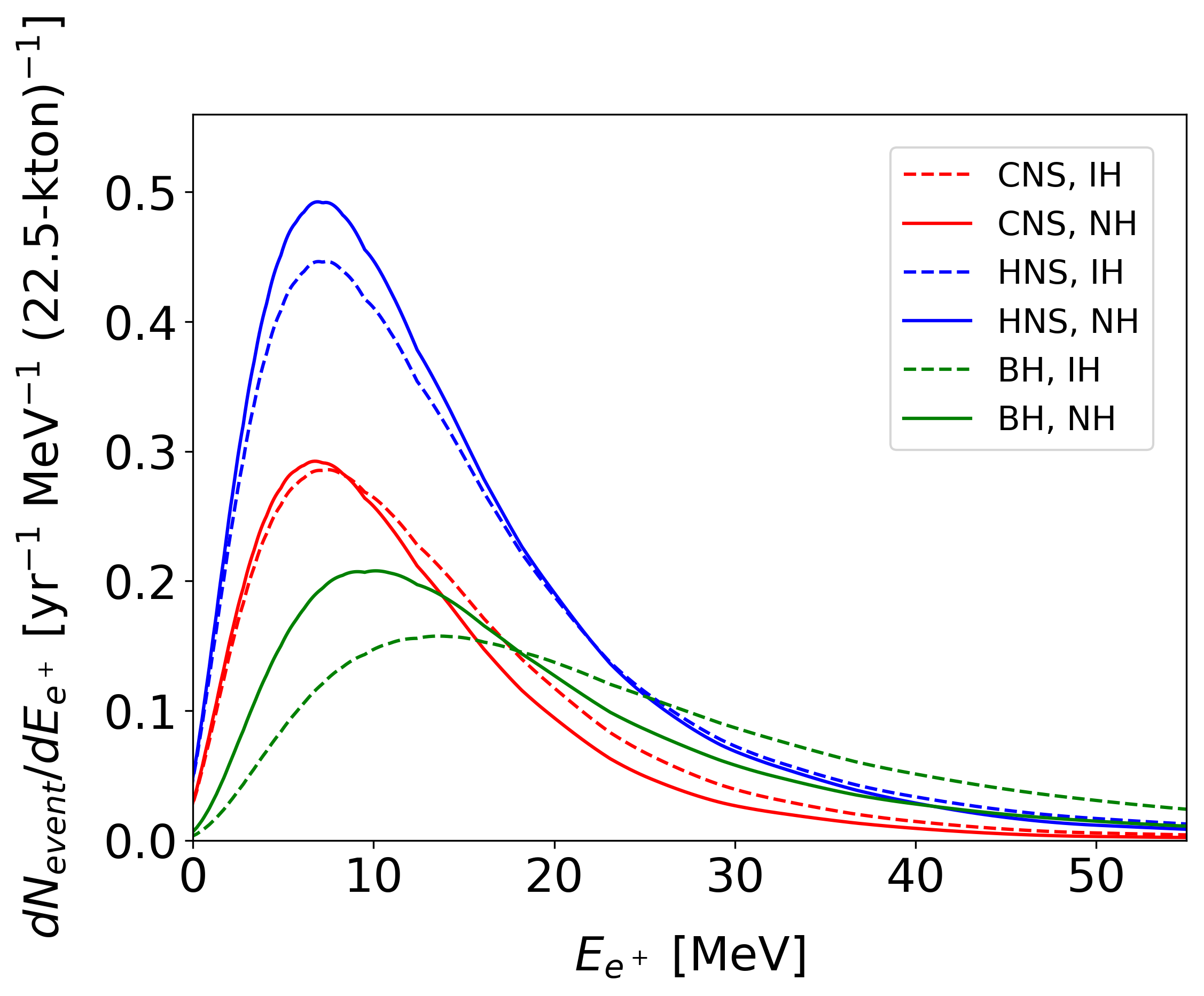}{0.33\textwidth}{(b) LS220 EOS}
            \fig{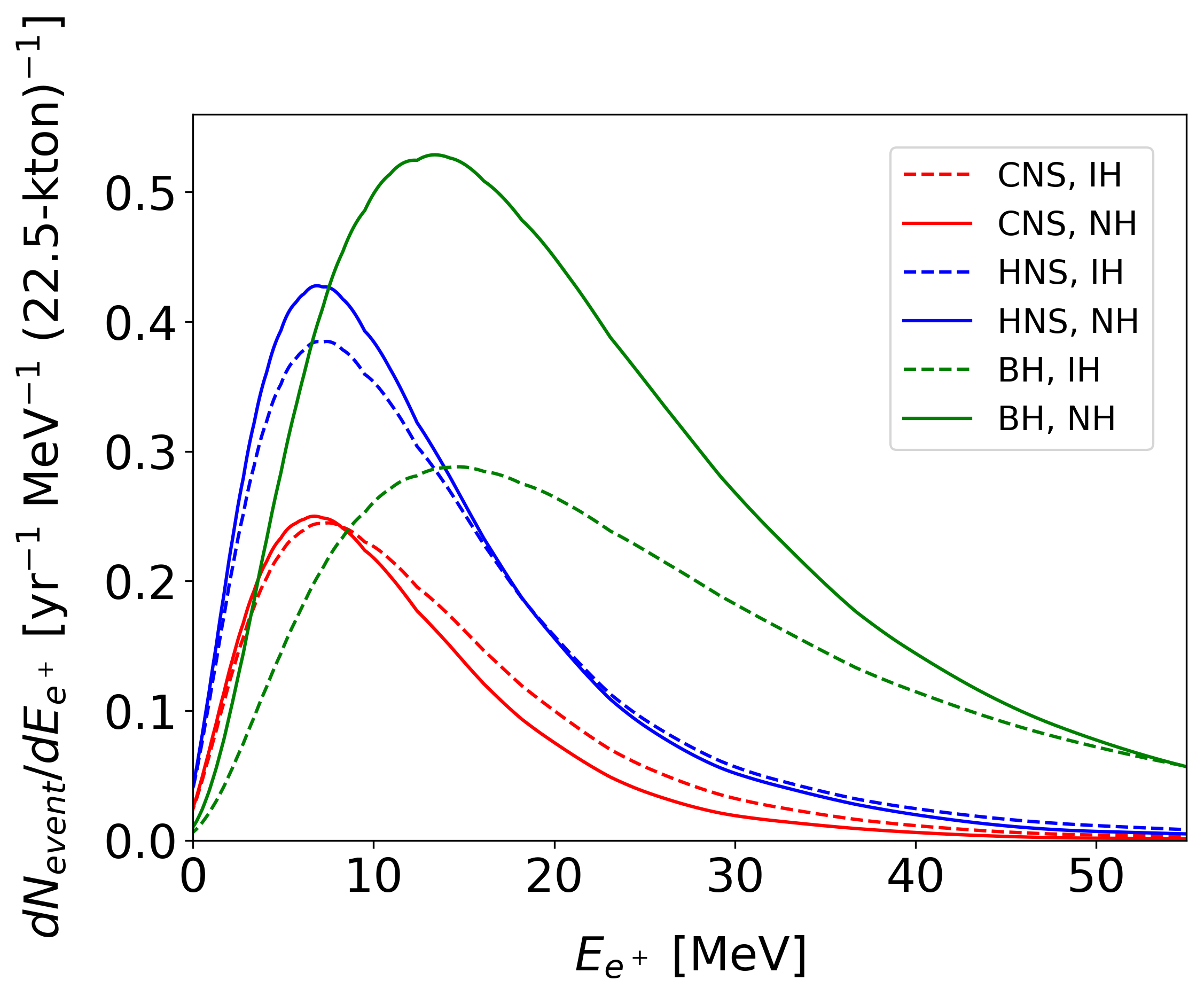}{0.33\textwidth}{(c) Shen EOS}}
  \vspace{+5truept}
  \caption{The event rate spectra of SRN $\bar\nu_e$ per 22.5~kton water per year obtained using the models of (a) Togashi EOS, (b) LS220 EOS, and (c) Shen EOS. The line and color conventions are the same as the ones used in Figure~\ref{fig:fluxcomp}.}
  \label{fig:neventcomp}
  \end{figure*}

The event rate spectra at an SK size water detector (22.5~kton) per year are shown in Figure~\ref{fig:neventcomp} for the single component models. If the same EOS and neutrino mass hierarchy are assumed, the SN models with a high-mass NS show a higher event rate than the SN models with a canonical-mass NS, while their normalized spectra look similar. This is again consistent with the strong NS mass dependence of the total neutrino energy and the weak NS mass dependence of the average neutrino energy as shown in Table~\ref{tab:snnmodel}. However, the contribution from the failed SN is not so simple. In the case of the Shen EOS models, the failed SN leads to a much higher SRN event rate than the SN with an NS. The Shen EOS has a higher maximum mass of NSs and a larger NS radius. Therefore, the amount of neutrinos emitted from a BH formation is larger and that from an NS formation is smaller. In contrast, the LS220 EOS models show the opposite trend; the failed SN model has a lower event rate due to the low maximum mass of NSs \citep[see Figure 13 of][]{2020ApJ...894....4D}. Whether the NS or BH leads to a higher SRN event rate depends on combination of the NS radius and maximum mass. As for the failed SN models, the difference between the cases with NH and IH is larger. This is because neutrinos are mainly emitted by the accretion of matter in the case of the failed SN and, thereby, $\bar\nu_e$ and $\nu_x$ have different spectra. In this process, the total energy is larger for $\bar\nu_e$ and the average energy is higher for $\nu_x$ (Table~\ref{tab:snnmodel}).

  %%%
  %%%

\subsection{Experimental sensitivity to the SRN flux} \label{sec:snstv}

\begin{figure*}[htbp]
  \gridline{\fig{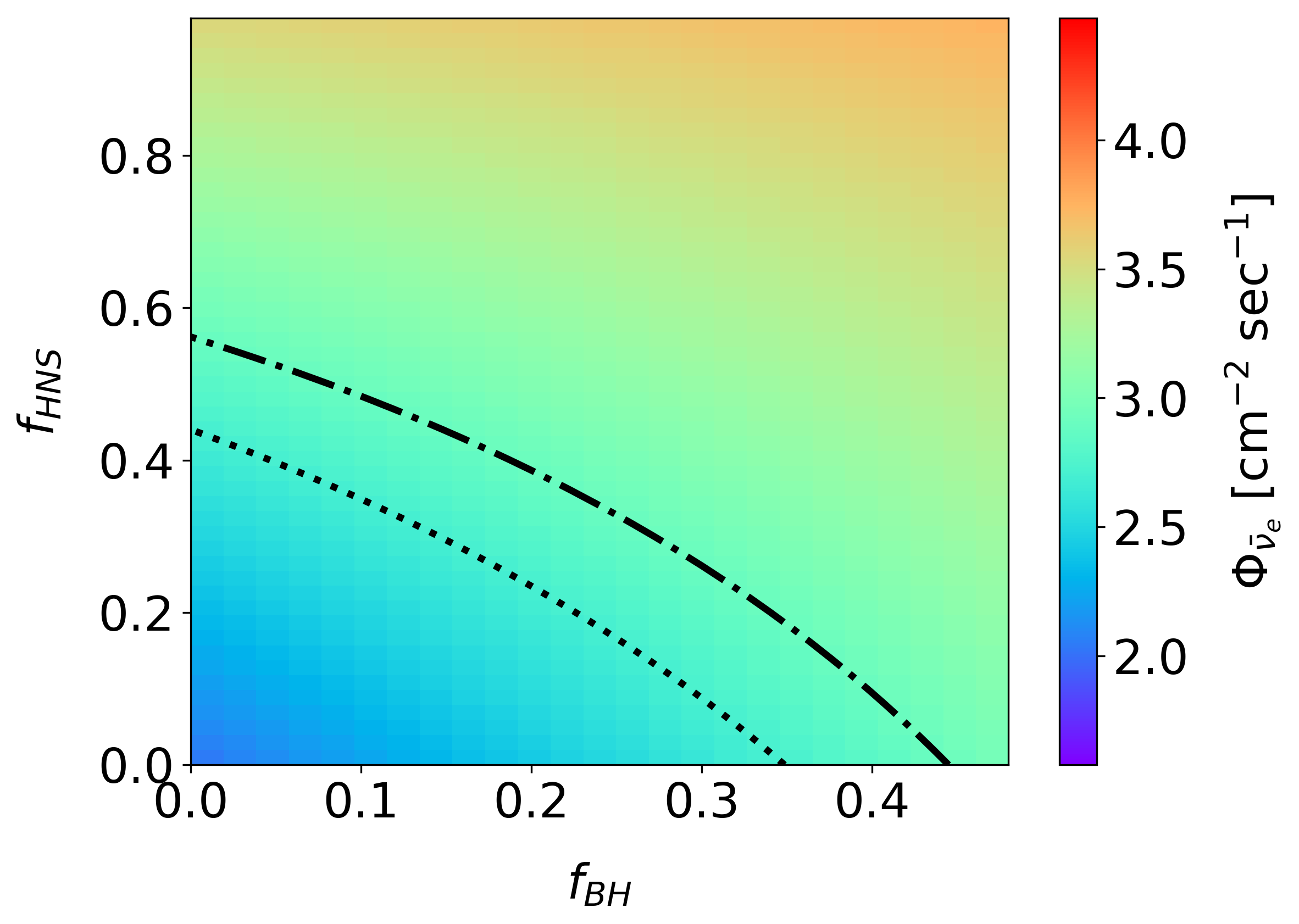}{0.37\textwidth}{(1a) Togashi EOS, \\ $13.3<E_\nu<31.3\,{\rm MeV}$}\hspace{-15truept}
            \fig{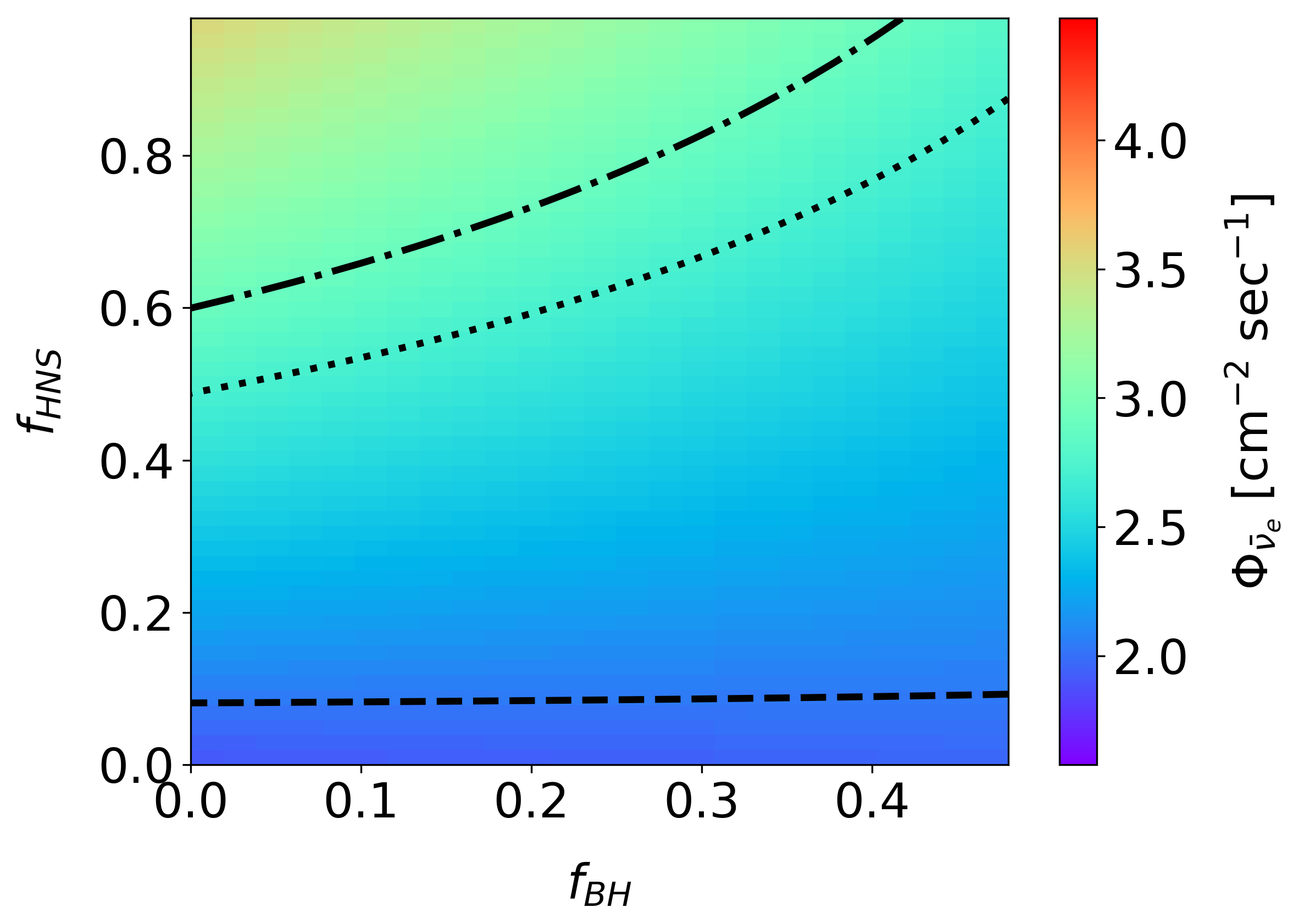}{0.37\textwidth}{(1b) LS220 EOS, \\ $13.3<E_\nu<31.3\,{\rm MeV}$}\hspace{-15truept}
            \fig{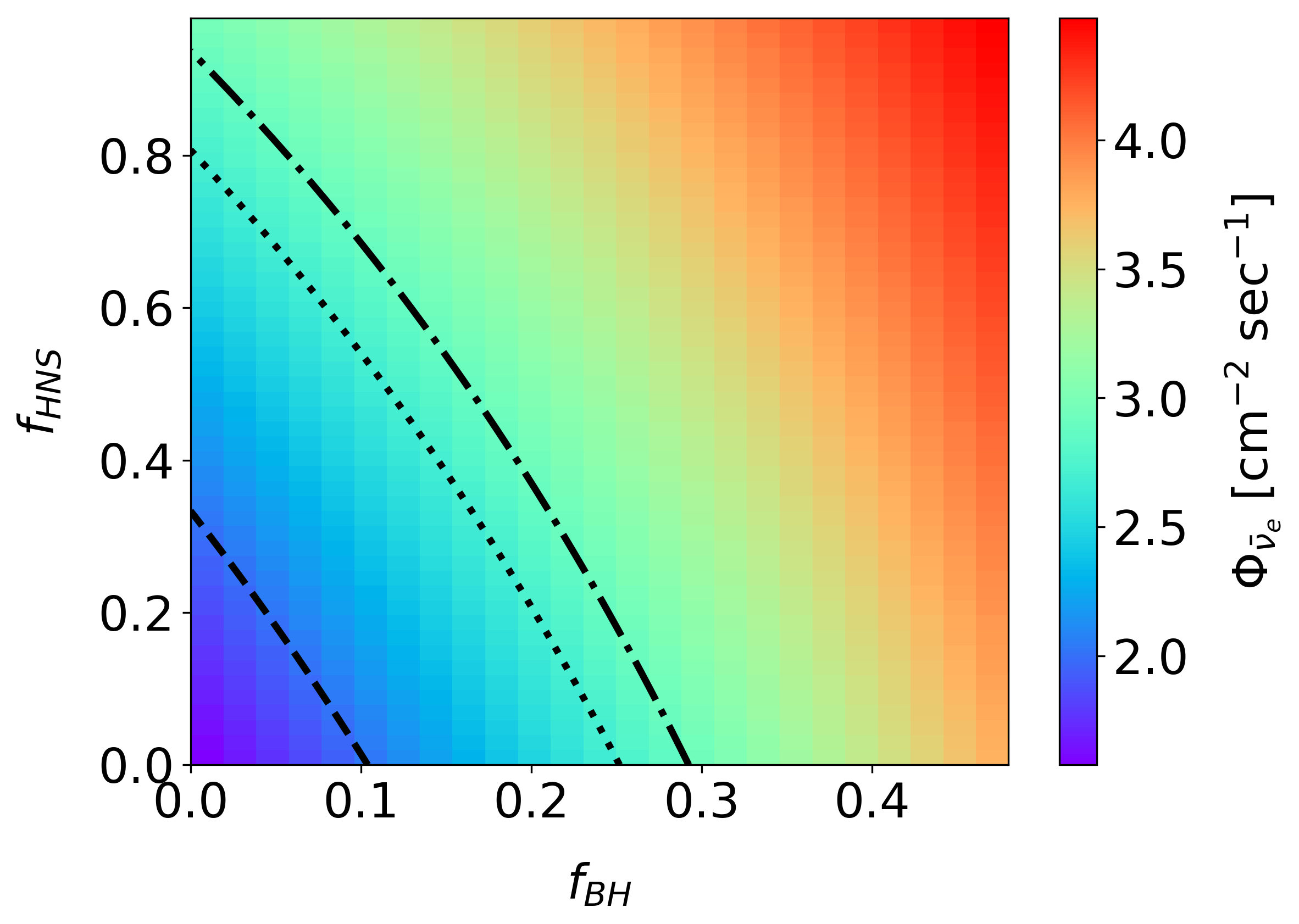}{0.37\textwidth}{(1c) Shen EOS, \\ $13.3<E_\nu<31.3\,{\rm MeV}$}}
  \gridline{\fig{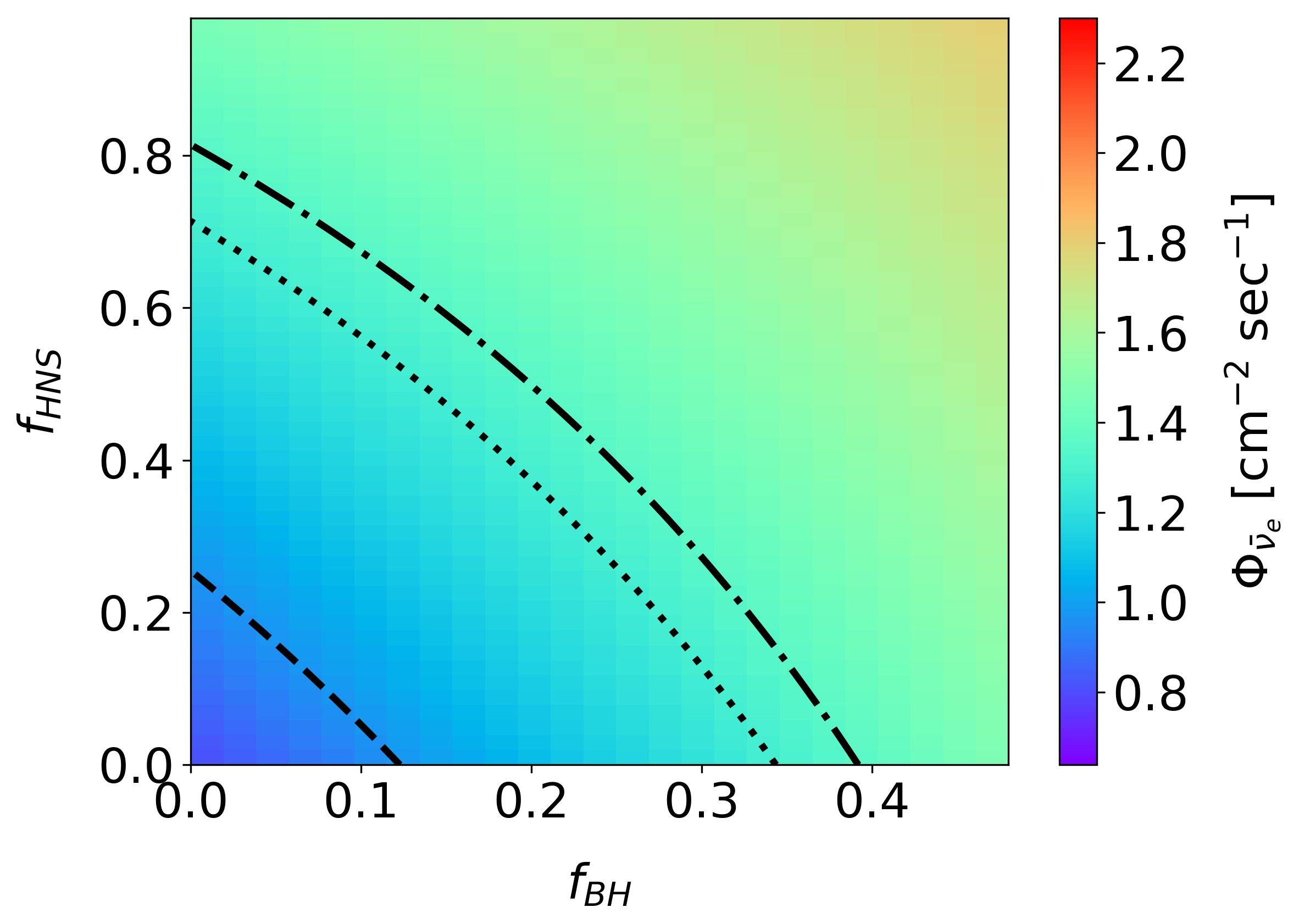}{0.37\textwidth}{(2a) Togashi EOS, \\ $17.3<E_\nu<31.3\,{\rm MeV}$}\hspace{-15truept}
            \fig{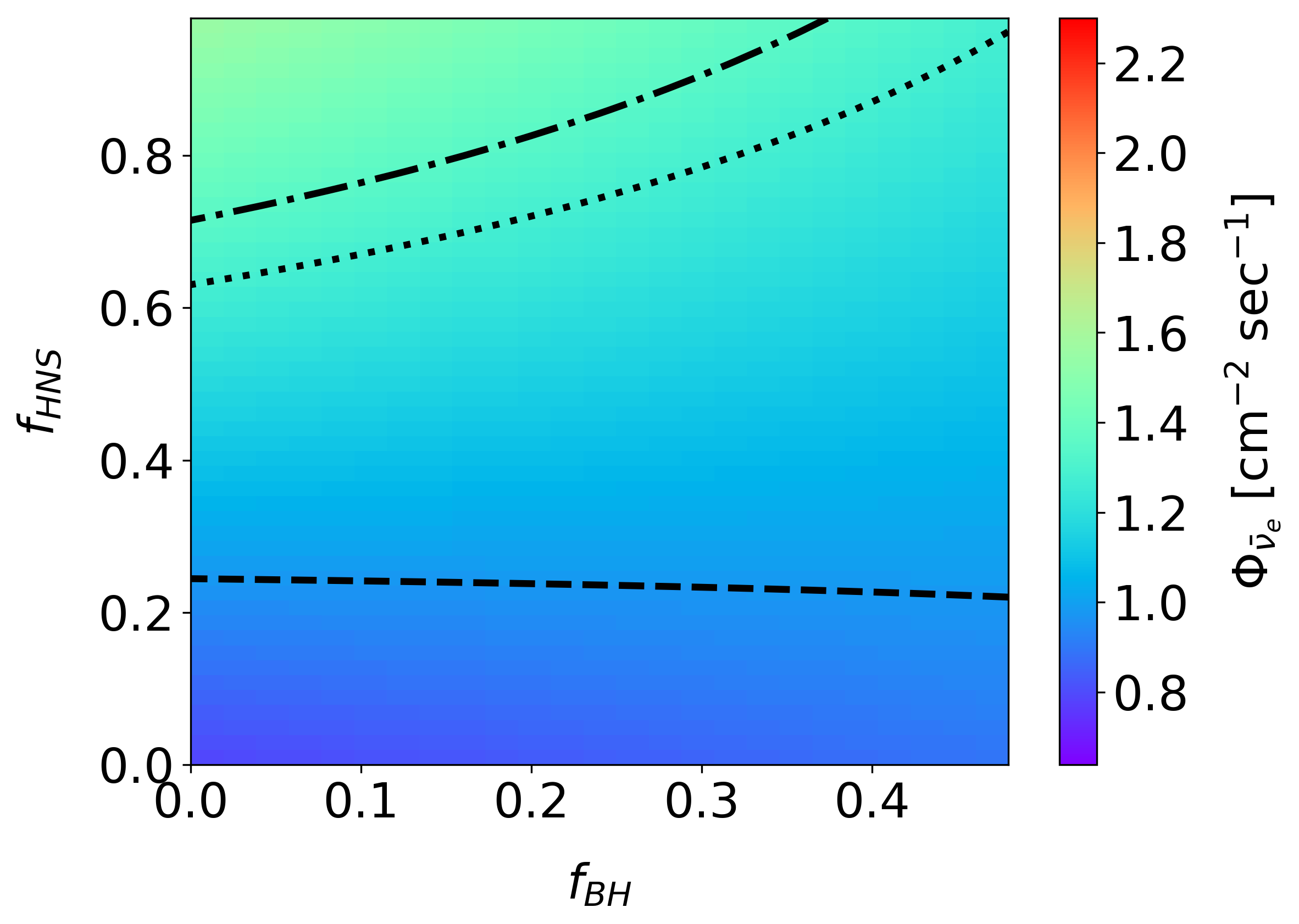}{0.37\textwidth}{(2b) LS220 EOS, \\ $17.3<E_\nu<31.3\,{\rm MeV}$}\hspace{-15truept}
            \fig{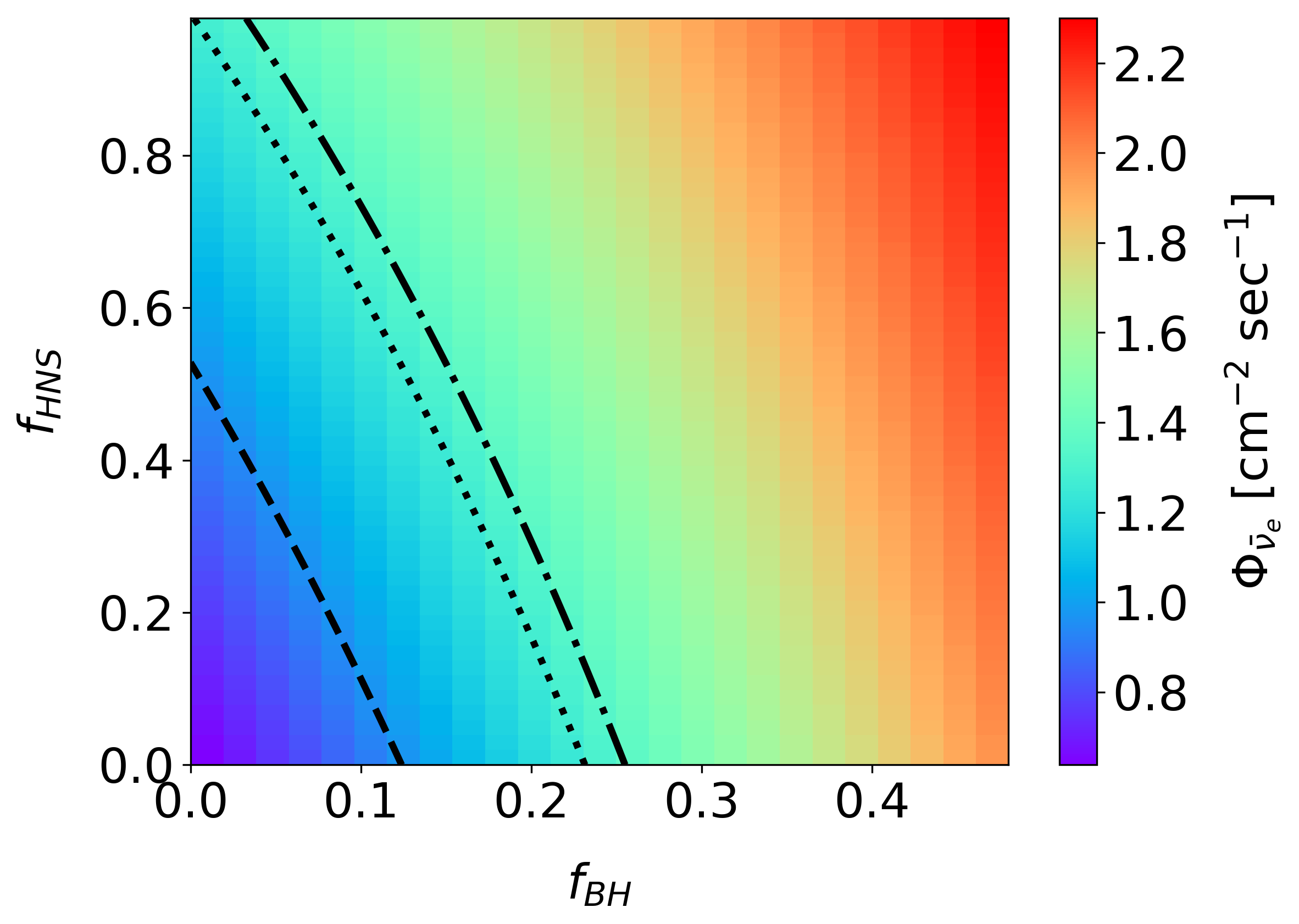}{0.37\textwidth}{(2c) Shen EOS, \\ $17.3<E_\nu<31.3\,{\rm MeV}$}}
  \vspace{+5truept}
  \caption{Two-dimensional map of the integrated SRN $\bar\nu_e$ flux as a function of $f_{\rm BH}$ and $f_{\rm HNS}$ for the cases with NH. Models with different EOSs and integration ranges of the neutrino energy, $E_\nu$ are shown in each panel: (1a) Togashi EOS and $13.3\,{\rm MeV}<E_\nu<31.3\,{\rm MeV}$, (1b) LS220 EOS and $13.3\,{\rm MeV}<E_\nu<31.3\,{\rm MeV}$, (1c) Shen EOS and $13.3\,{\rm MeV}<E_\nu<31.3\,{\rm MeV}$, (2a) Togashi EOS and $17.3\,{\rm MeV}<E_\nu<31.3\,{\rm MeV}$, (2b) LS220 EOS and $17.3\,{\rm MeV}<E_\nu<31.3\,{\rm MeV}$, and (2c) Shen EOS and $17.3\,{\rm MeV}<E_\nu<31.3\,{\rm MeV}$. Dashed-dotted, dotted, and dashed lines represent the $2\sigma$ C.L. expected sensitivities at SK-Gd over 10 yr, HK over 3 yr, and HK over 5 yr, respectively. HK over 5 yr could test the entire parameter space at $2\sigma$ C.L. for the (1a) case.}
  \label{fig:snstv2dallnh}
  \end{figure*}

\begin{figure*}[htbp]
  \gridline{\fig{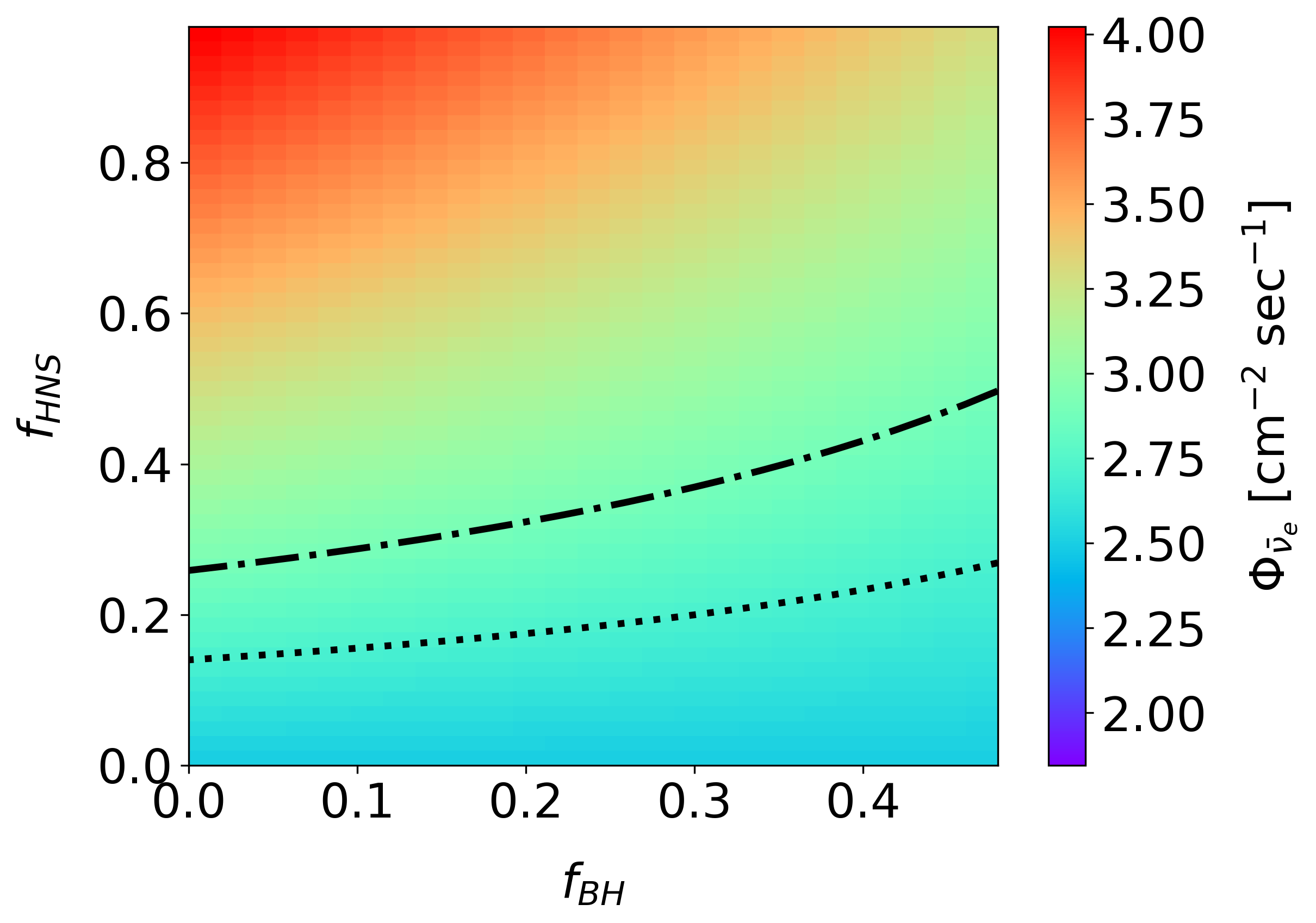}{0.37\textwidth}{(1a) Togashi EOS, \\ $13.3<E_\nu<31.3\,{\rm MeV}$}\hspace{-15truept}
            \fig{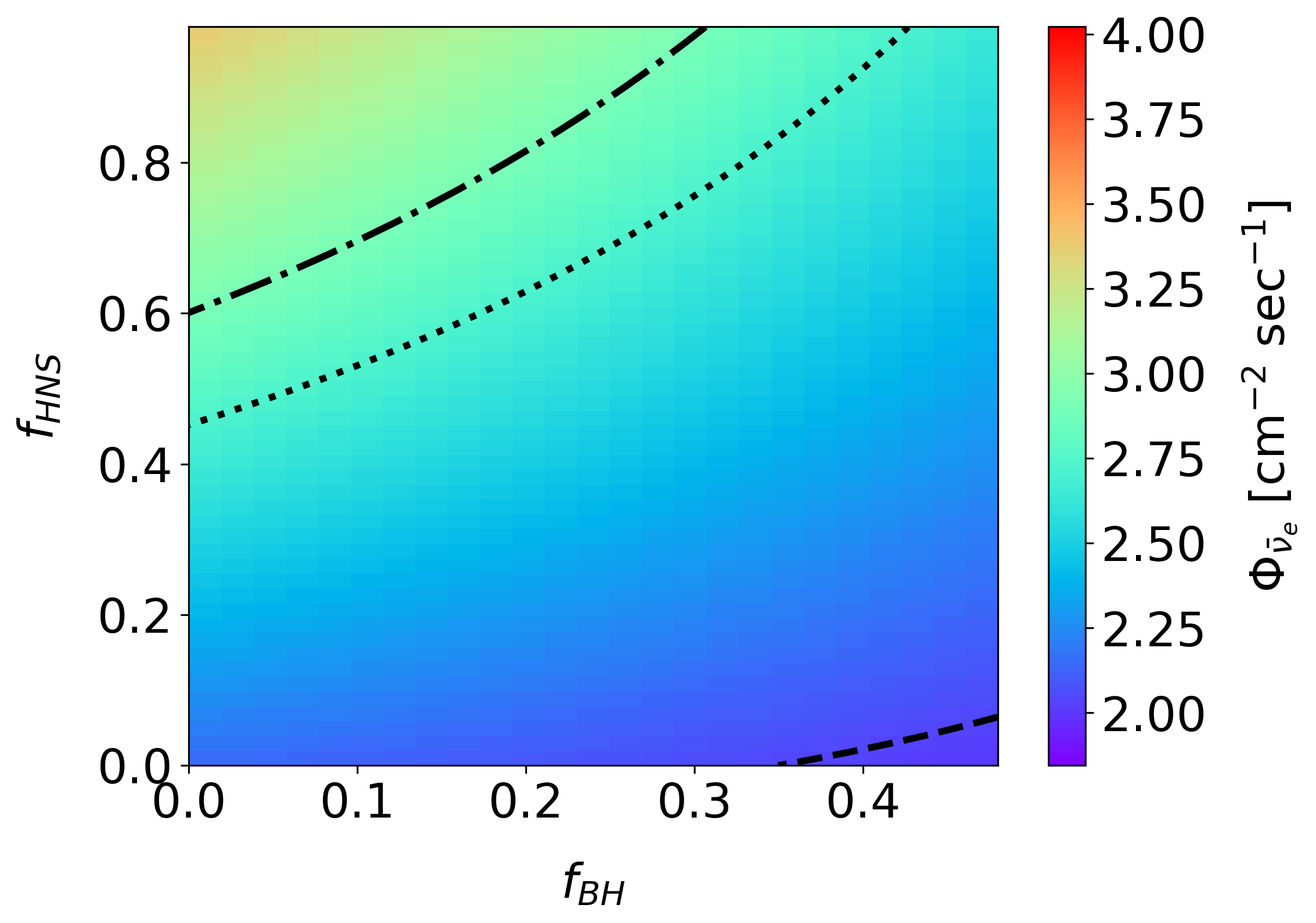}{0.37\textwidth}{(1b) LS220 EOS, \\ $13.3<E_\nu<31.3\,{\rm MeV}$}\hspace{-15truept}
            \fig{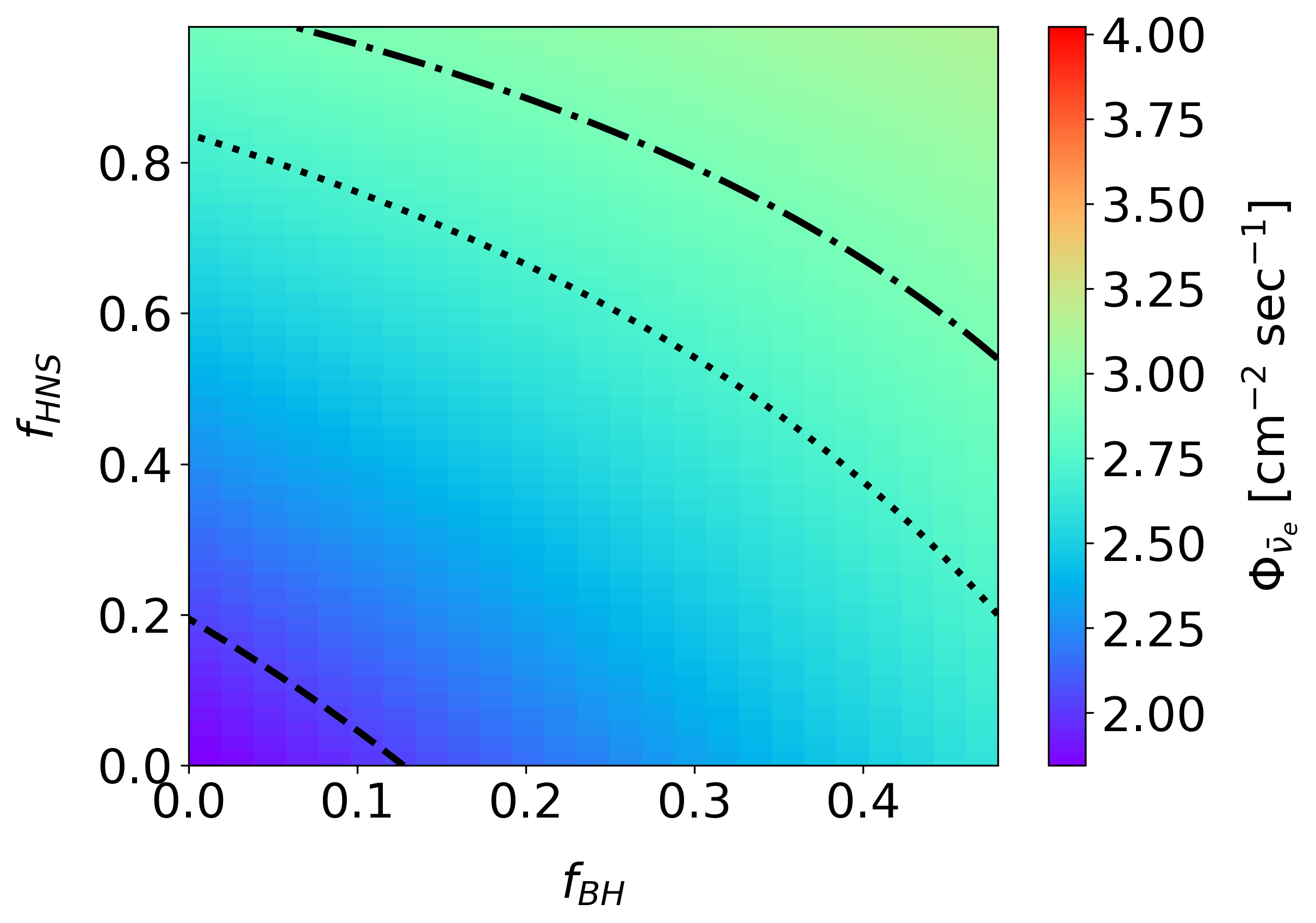}{0.37\textwidth}{(1c) Shen EOS, \\ $13.3<E_\nu<31.3\,{\rm MeV}$}}
  \gridline{\fig{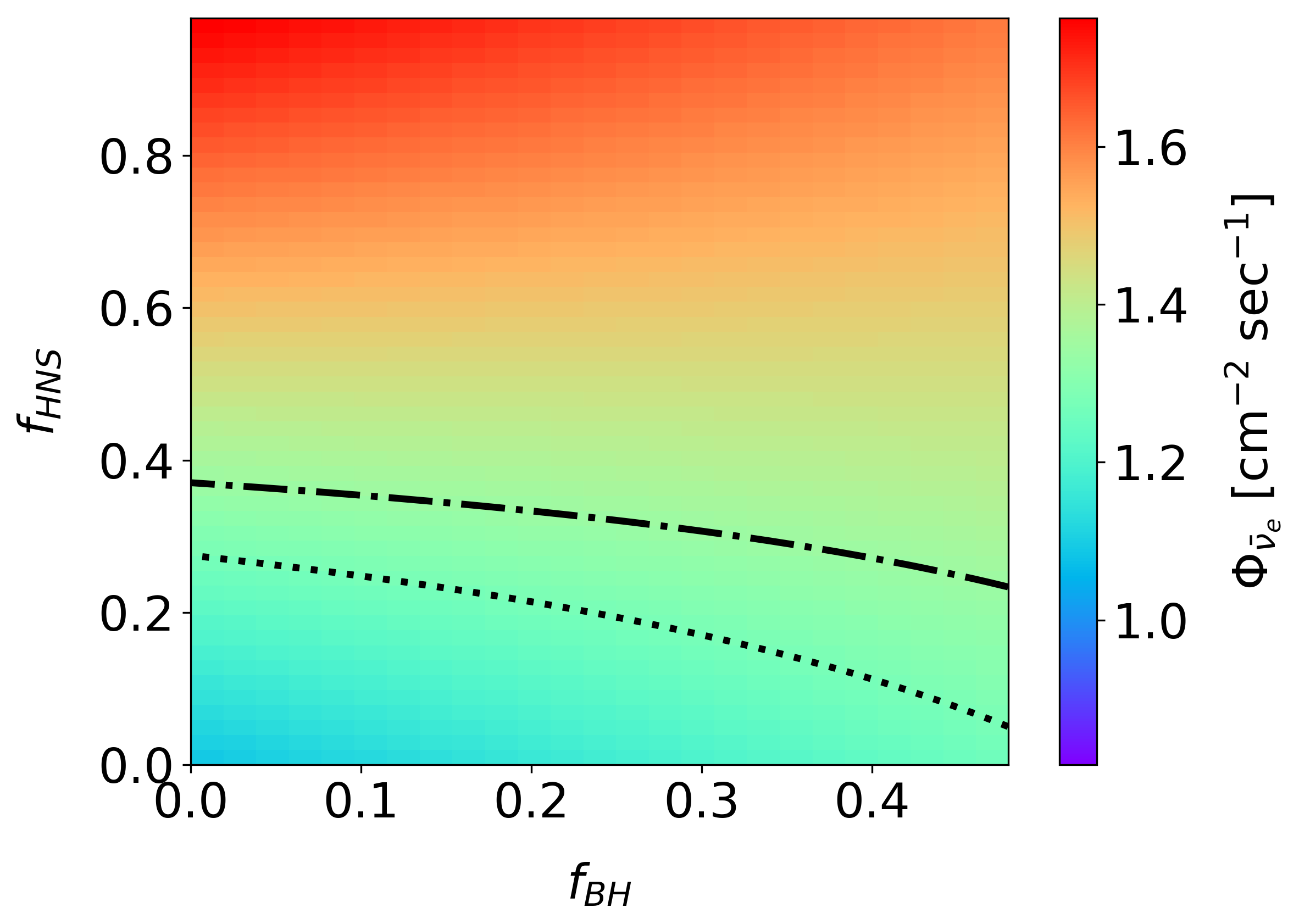}{0.37\textwidth}{(2a) Togashi EOS, \\ $17.3<E_\nu<31.3\,{\rm MeV}$}\hspace{-15truept}
            \fig{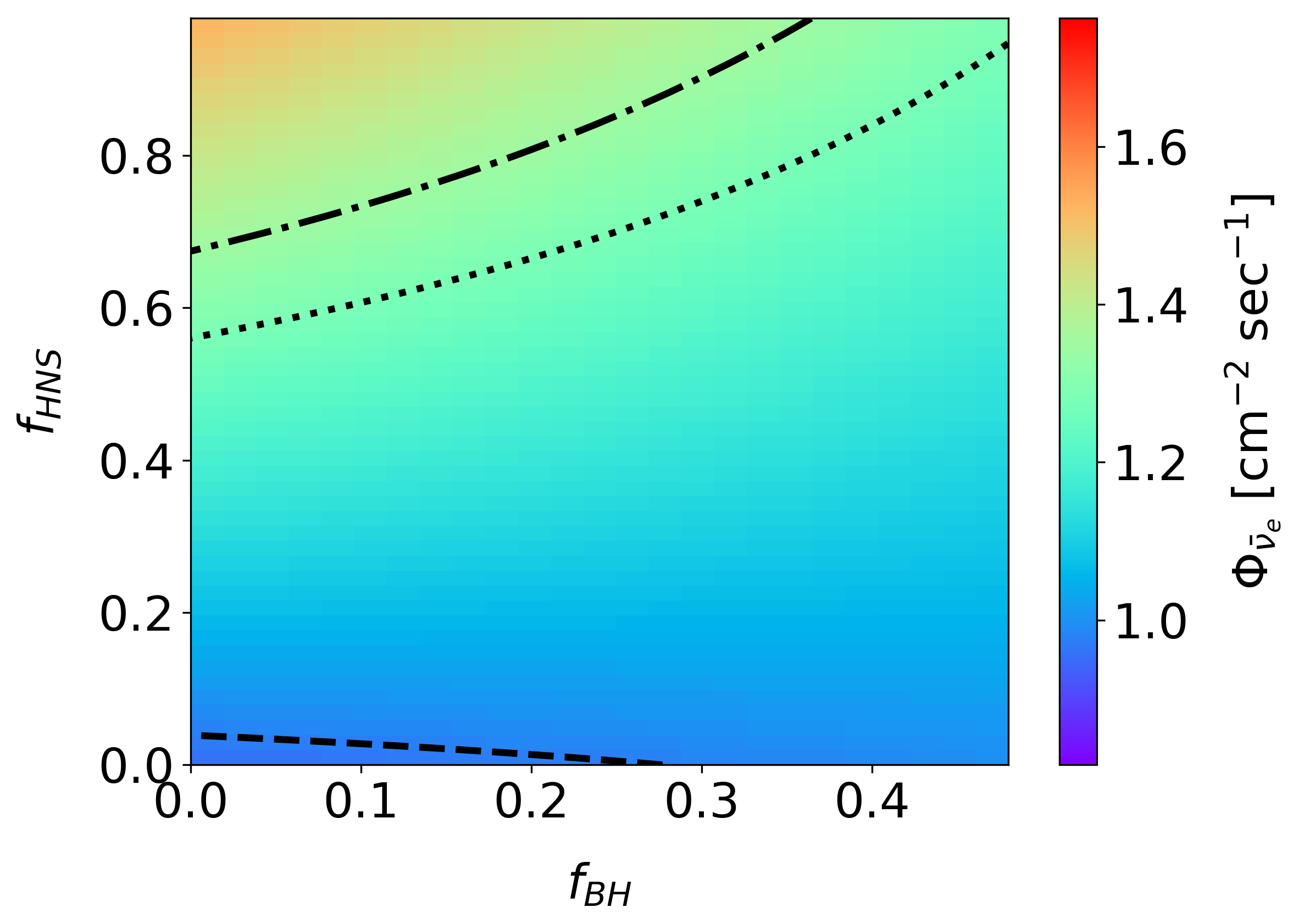}{0.37\textwidth}{(2b) LS220 EOS, \\ $17.3<E_\nu<31.3\,{\rm MeV}$}\hspace{-15truept}
            \fig{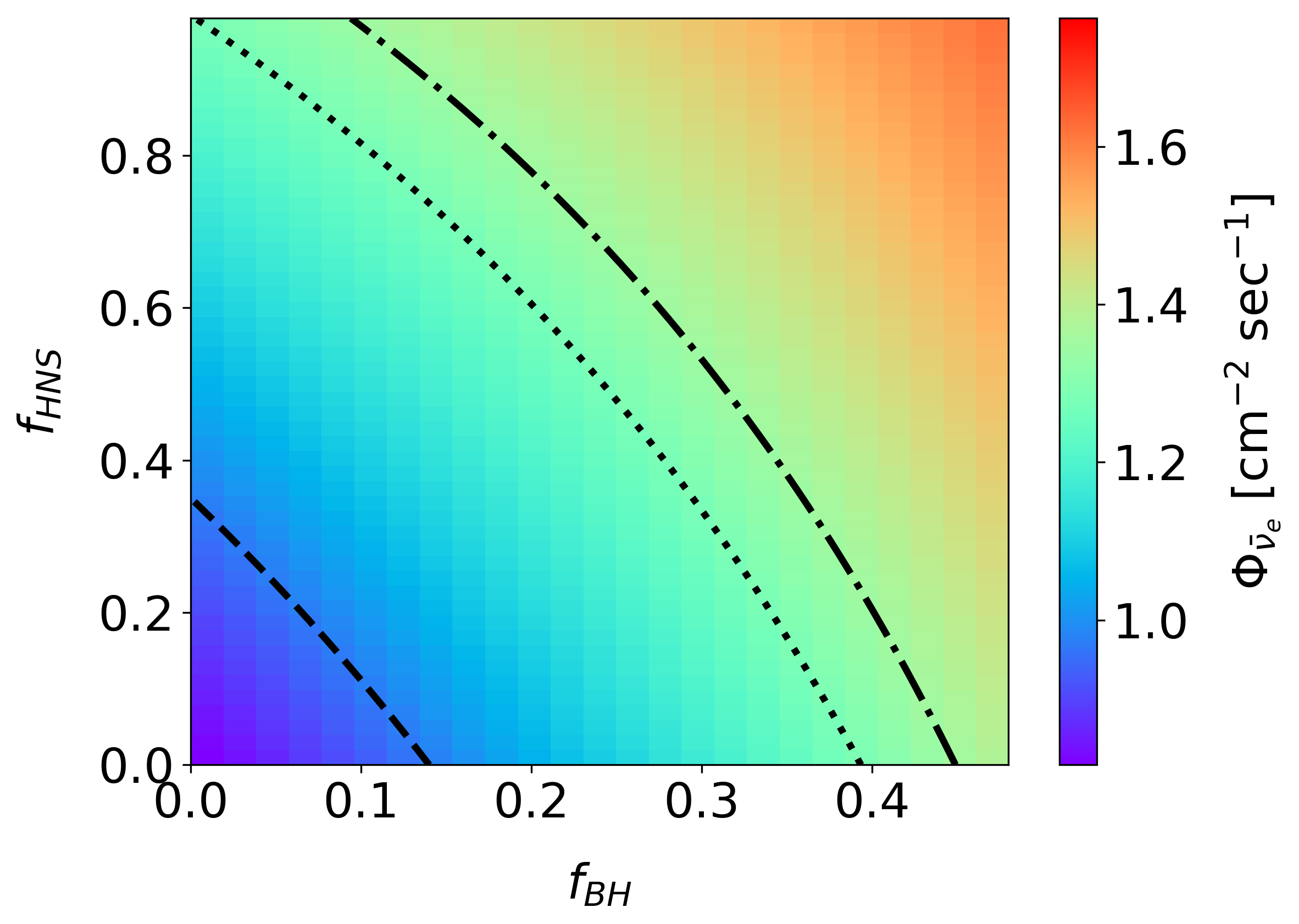}{0.37\textwidth}{(2c) Shen EOS, \\ $17.3<E_\nu<31.3\,{\rm MeV}$}}
  \vspace{+5truept}
  \caption{Same distributions with the $2\sigma$ C.L. expected sensitivity lines as the ones in Figure~\ref{fig:snstv2dallnh} but for the cases with IH. HK over 5 yr could test the entire parameter space at $2\sigma$ C.L. for the Togashi EOS cases in (1a) and (2a).}
  \label{fig:snstv2dallih}
  \end{figure*}

Here we combine the contributions of all three components. Then, dependence on $f_{\rm BH}$ and $f_{\rm HNS}$ is investigated with the sensitivity study. In Figure~\ref{fig:snstv2dallnh}, the $2\sigma$ C.L. expected sensitivities on the $\bar\nu_e$ flux at SK-Gd over 10 yr, HK over 3 yr, and HK over 5 yr are shown on top of the integrated SRN $\bar\nu_e$ flux for different energy regions, (1) $13.3<E_\nu<31.3\,{\rm MeV}$ and (2) $17.3<E_\nu<31.3\,{\rm MeV}$, with NH. Figure~\ref{fig:snstv2dallih} shows the corresponding distributions with IH. We can recognize that the sensitivity is improved by lowering the energy threshold to 13.3~MeV in some cases. This is thanks to the latest extensive study on the background in the low energy regime as described in Section~\ref{sec:bkg}. Nevertheless, for some parameter sets, the sensitivity is better for $17.3<E_\nu<31.3\,{\rm MeV}$. This is likely for the cases with a large $f_{\rm BH}$ because the failed SN models have relatively large fluxes in $E_\nu>17.3\,{\rm MeV}$ due to their hard spectra. Therefore, in the following, we plot the projected sensitivity by combining the sensitivities for $13.3<E_\nu<31.3\,{\rm MeV}$ and $17.3<E_\nu<31.3\,{\rm MeV}$.

  %%%
  %%%

  %%%
  %%%

\begin{figure*}[htbp]
    \centering
    \begin{minipage}{0.45\hsize}
        \begin{center}
            \includegraphics[clip,width=7.0cm]{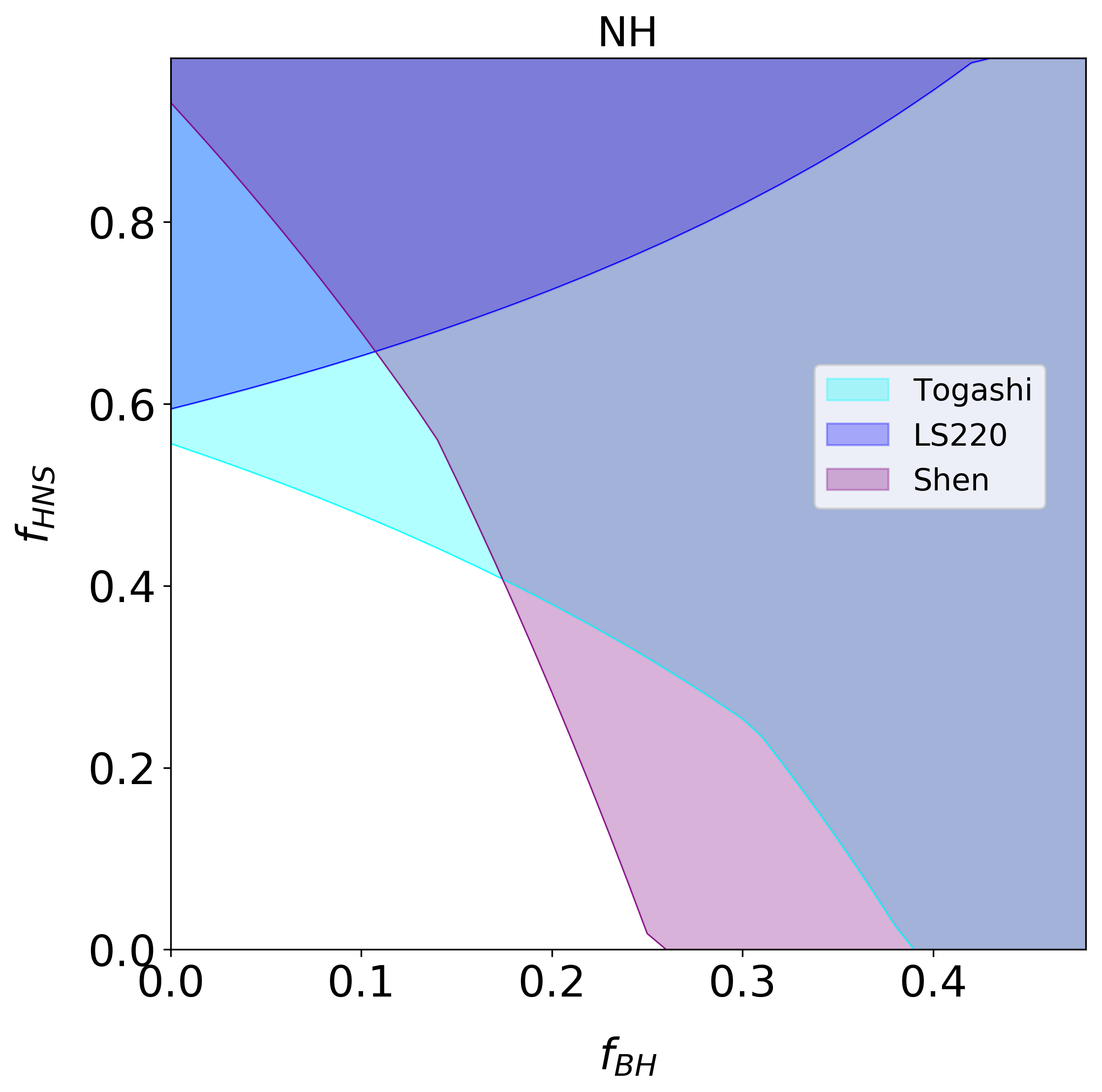}
        \end{center}
    \end{minipage}
    \begin{minipage}{0.45\hsize}
        \begin{center}
            \includegraphics[clip,width=7.0cm]{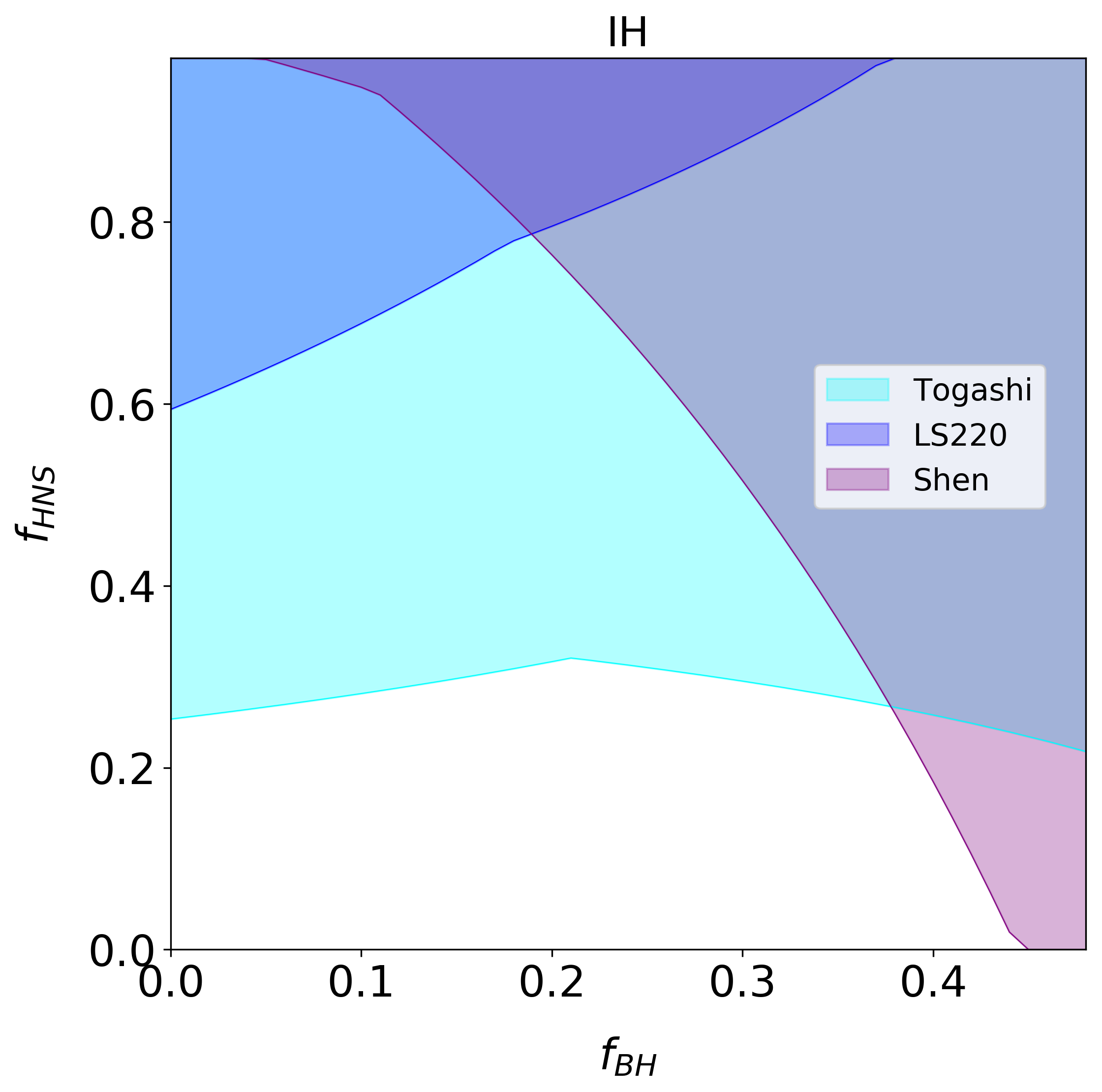}
        \end{center}
    \end{minipage}
    \vspace{+10truept}
    \caption{Detectable regions on the $f_{\rm BH}$--$f_{\rm HNS}$ plane at SK-Gd over 10 yr for three different EOSs for NH (left) and IH (right). The colored areas represent the regions excluded from either $13.3\,{\rm MeV}<E_\nu<31.3\,{\rm MeV}$ or $17.3\,{\rm MeV}<E_\nu<31.3\,{\rm MeV}$ at $2\sigma$ C.L. Note that these are not statistically combined regions.}
    \label{fig:cntr2derange}
  \end{figure*}

In Figure~\ref{fig:cntr2derange}, the detectable regions at SK-Gd over 10 yr are shown for different EOS models. We find that the region at a higher $f_{\rm HNS}$ is more likely to be detected. This is because a larger amount of neutrinos are emitted from a high-mass NS than a canonical-mass NS for every EOS model while their normalized SRN spectra look similar (Figure~\ref{fig:neventcomp}). In contrast, dependence on $f_{\rm BH}$ is different among the EOS models. Since the SRN flux of the failed SN with a BH is lower than that of the SN with a high-mass NS for the case of the LS220 EOS (Figure~\ref{fig:integcomp}), the event rate decreases with increasing $f_{\rm BH}$. It should be emphasized that inclusion of failed SNe does not always enhance the possibility of the SRN detection; it depends on the EOS. In any case, whether SK-Gd can reach the $2\sigma$ level over 10 yr depends on $f_{\rm BH}$ and $f_{\rm HNS}$. Thus, within 10 yr, SK-Gd would provide some implications about the distribution of compact remnants formed by stellar core collapse.

\begin{figure*}[htbp]
  \gridline{\fig{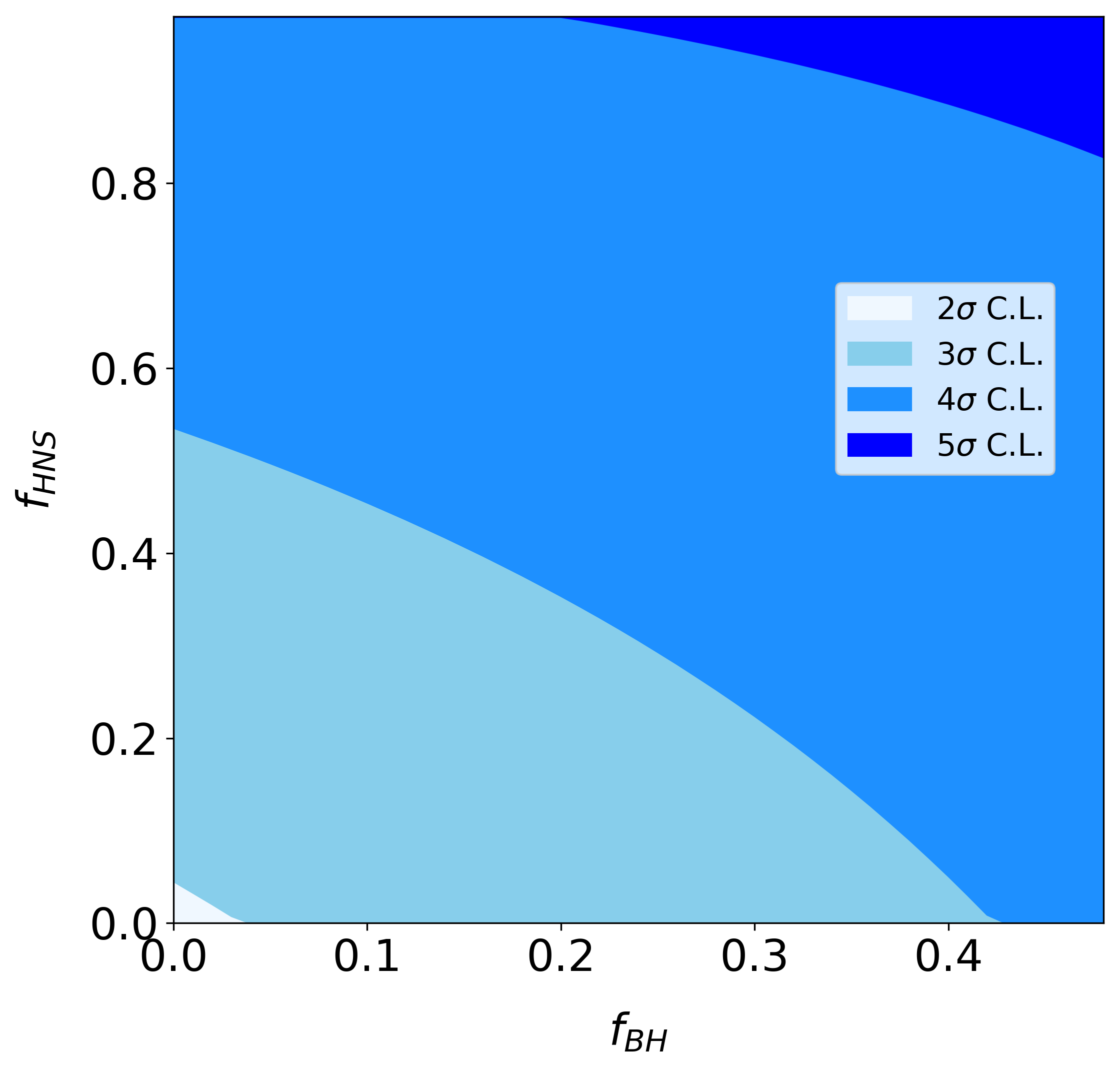}{0.32\textwidth}{(1a) Togashi EOS, NH}\hspace{-15truept}
            \fig{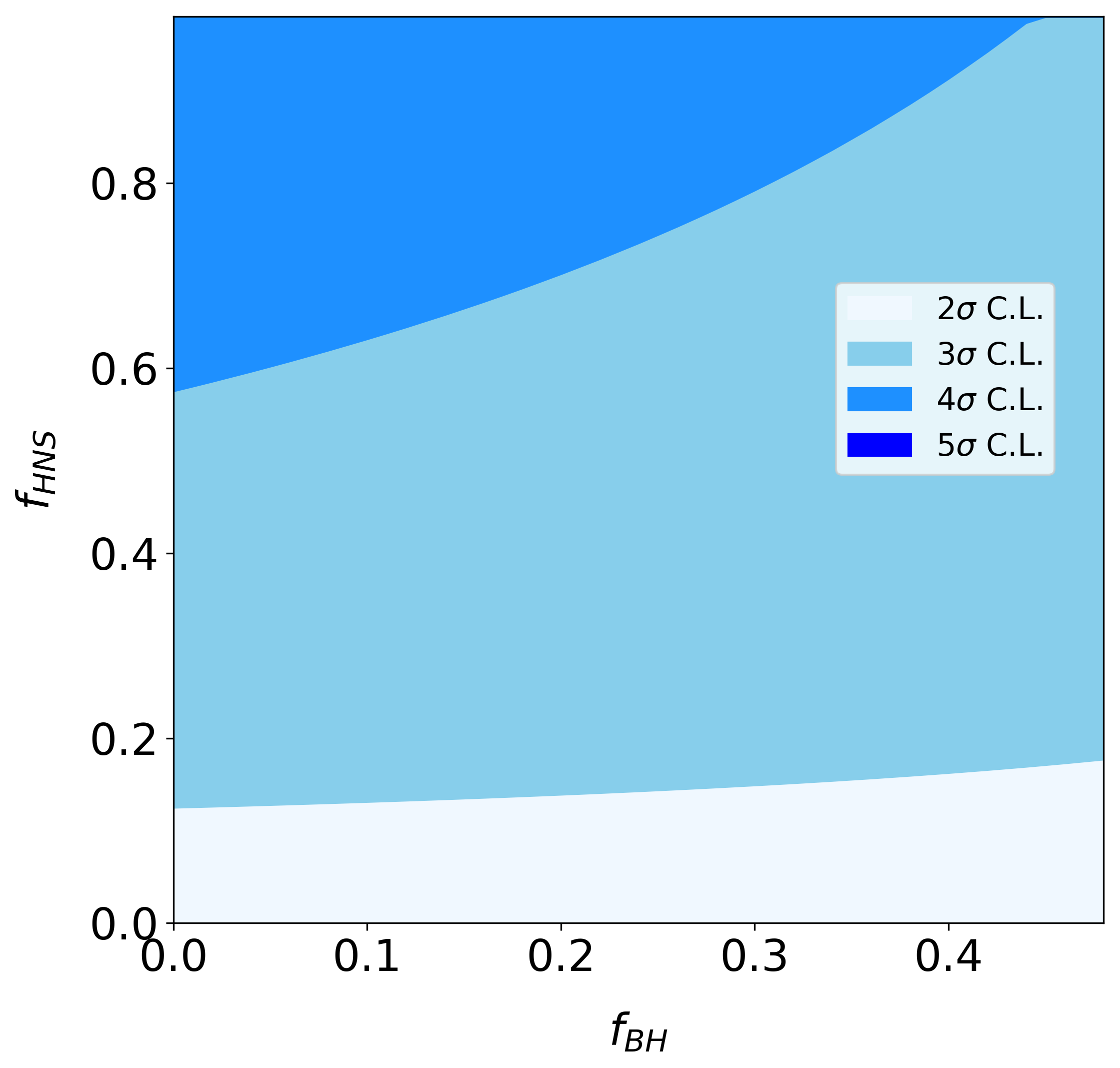}{0.32\textwidth}{(1b) LS220 EOS, NH}\hspace{-15truept}
            \fig{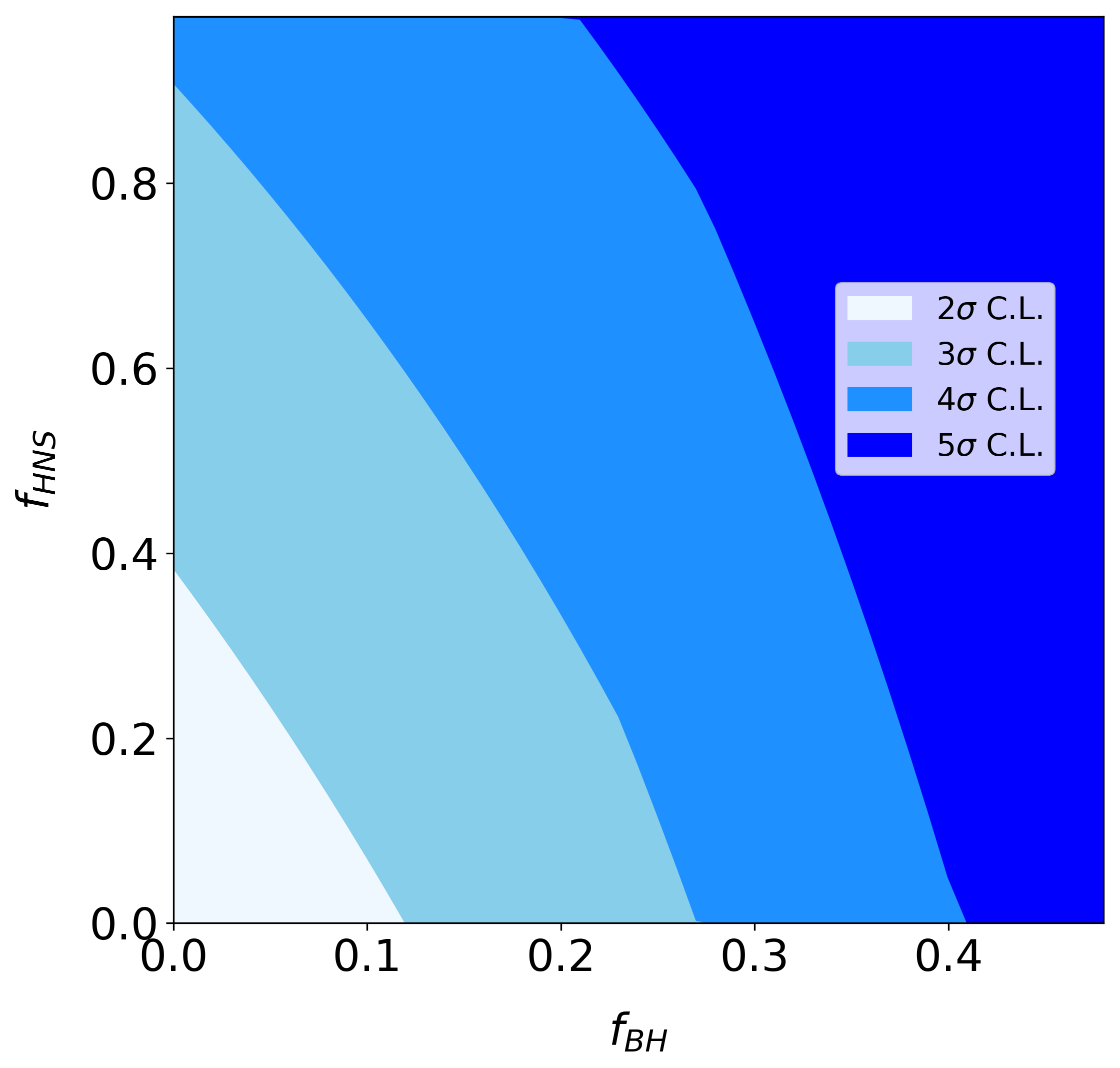}{0.32\textwidth}{(1c) Shen EOS, NH}}
  \gridline{\fig{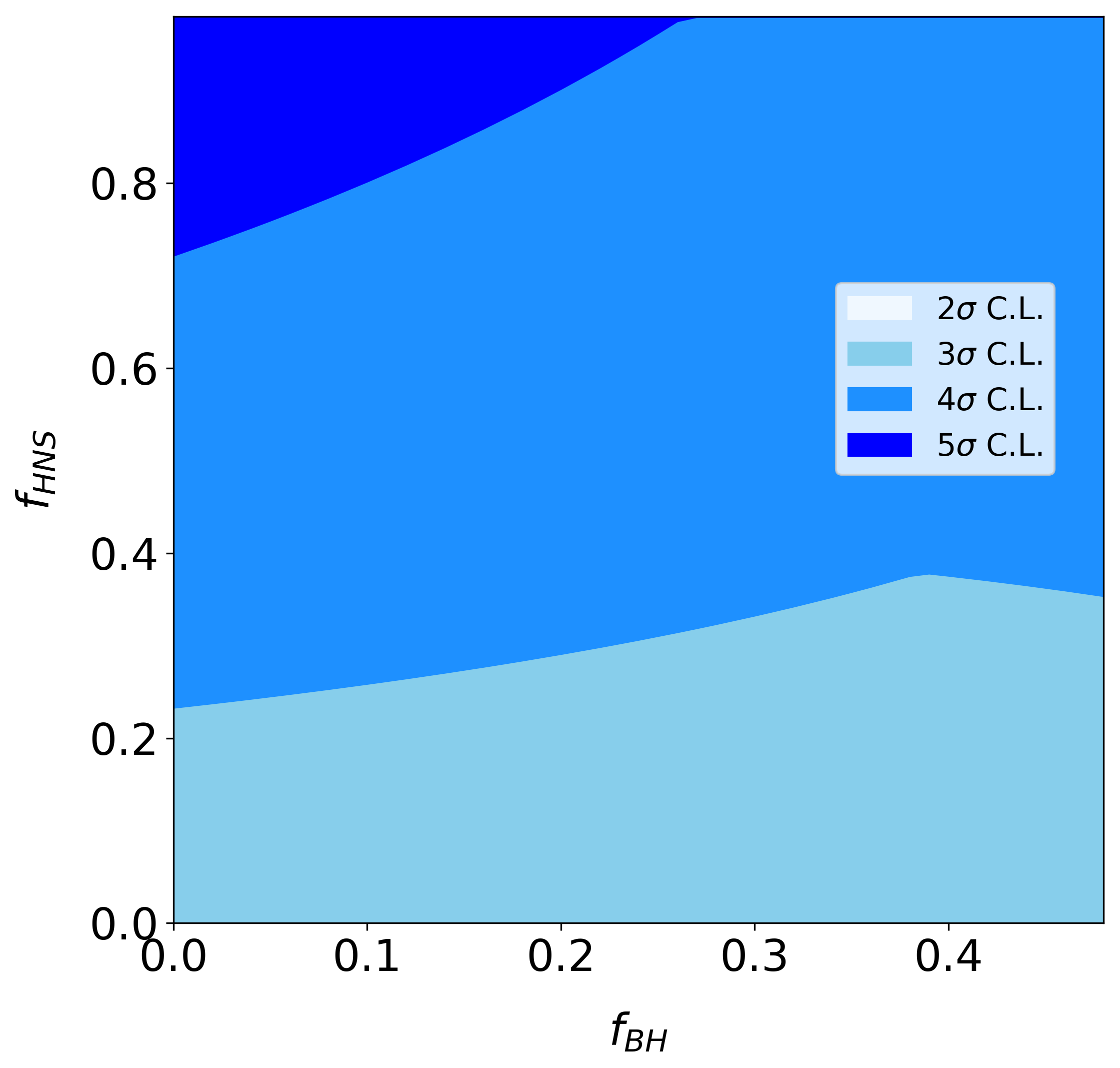}{0.32\textwidth}{(2a) Togashi EOS, IH}\hspace{-15truept}
            \fig{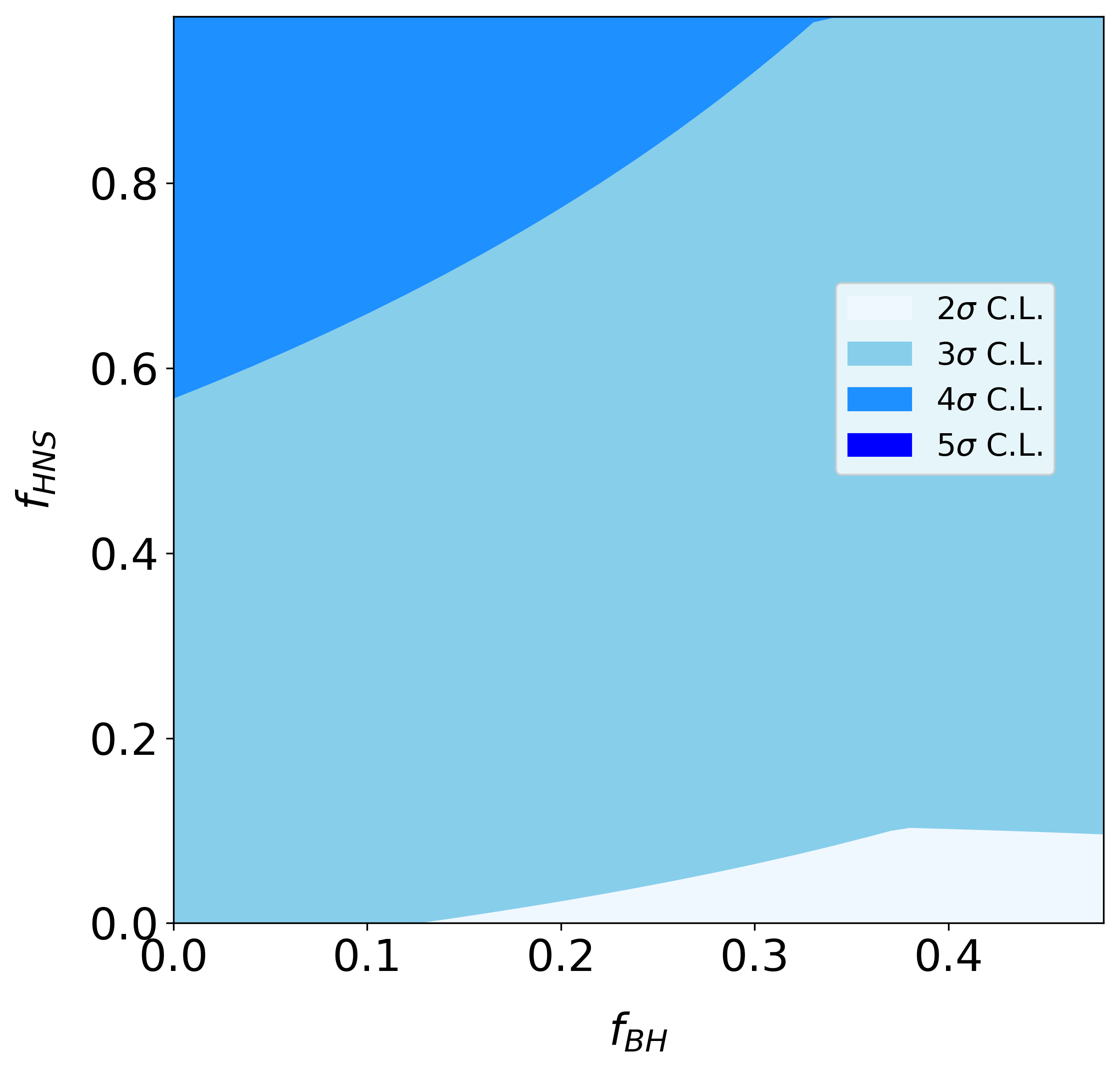}{0.32\textwidth}{(2b) LS220 EOS, IH}\hspace{-15truept}
            \fig{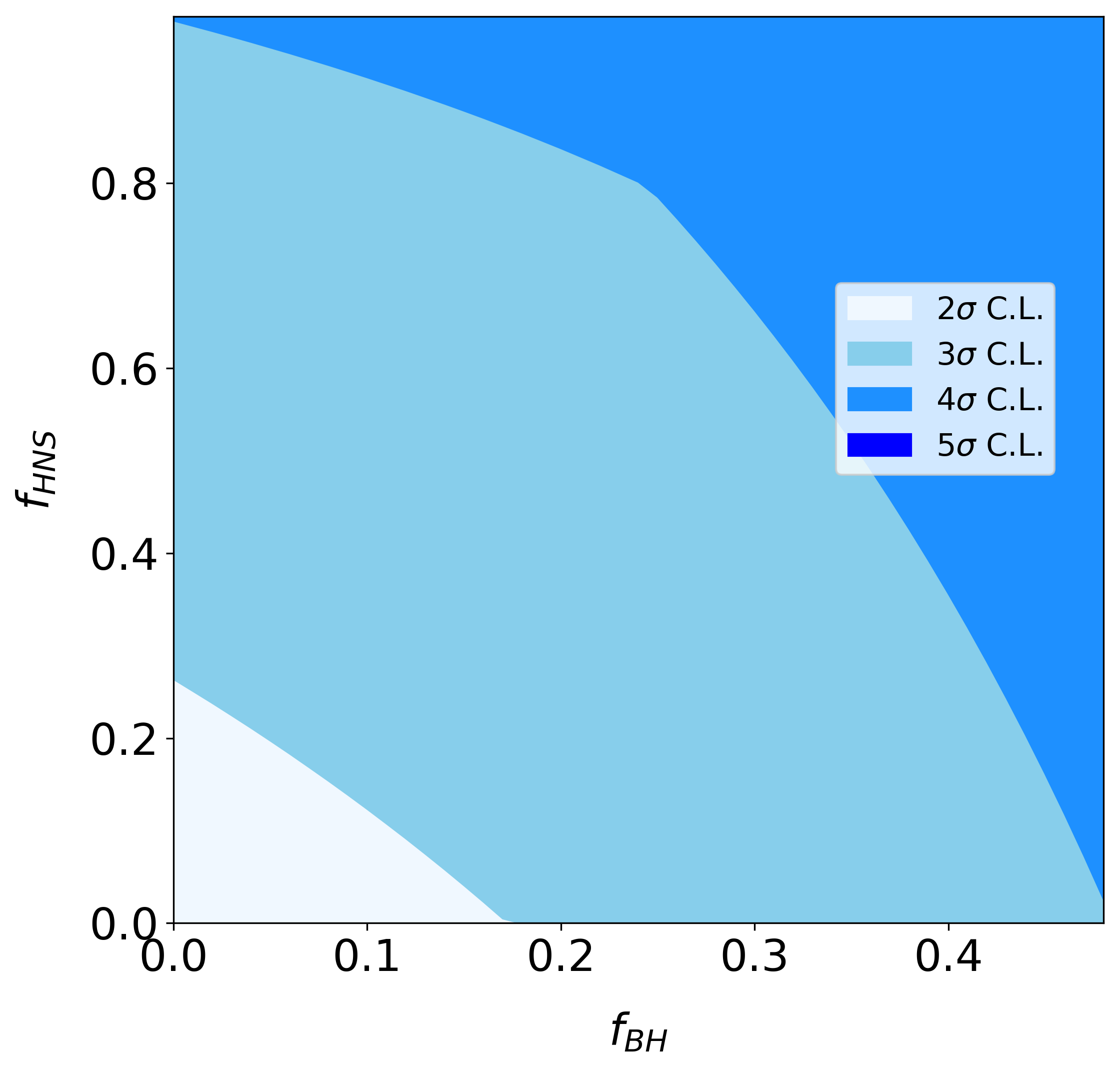}{0.32\textwidth}{(2c) Shen EOS, IH}}
  \vspace{+5truept}
  \caption{Same distributions as Figure~\ref{fig:cntr2derange} but for HK 10 yr and different C.L. Here the results with three different EOSs are shown for NH (top) and IH (bottom).}
  \label{fig:cntr2dyear}
  \end{figure*}

We now turn to the SRN observation with HK. Figure~\ref{fig:cntr2dyear} shows the detectable regions at HK over 10 yr for different C.L., EOS models, and neutrino mass hierarchies. The results show that the whole parameter space could be tested at a $2\sigma$ level for all cases and a large region of the parameter space is covered at a $3\sigma$ level. A higher significance could be achieved for only some cases. Note that our assumptions for the HK study still contains a lot more uncertain factors than those for SK-Gd. 

  %%%
  %%%
  
  %%%
  %%%

\subsection{Discussion} \label{sec:discuss}

\begin{figure*}[htbp]
  \gridline{\fig{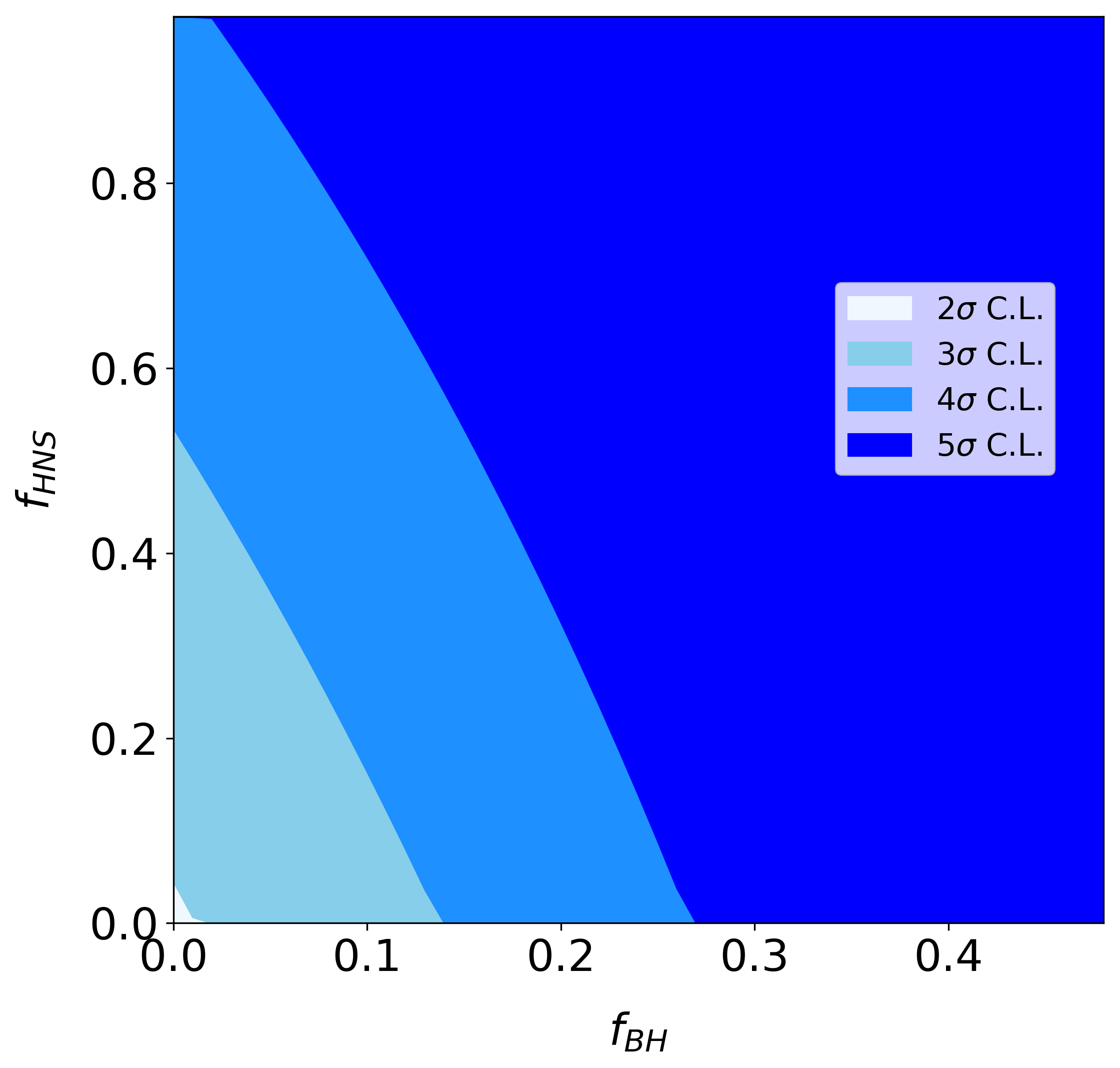}{0.32\textwidth}{(1a) Togashi EOS, NH}\hspace{-15truept}
            \fig{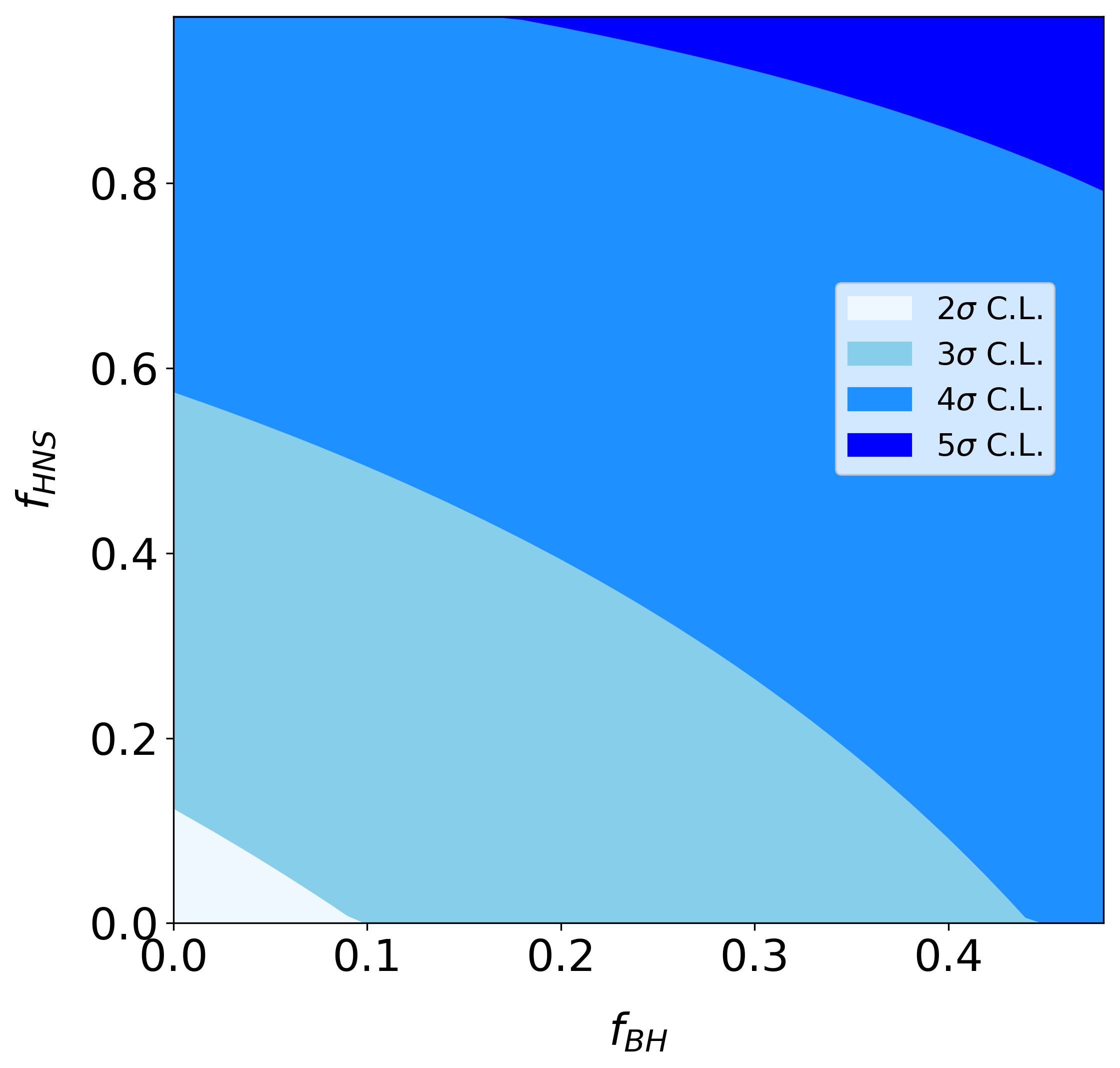}{0.32\textwidth}{(1b) LS220 EOS, NH}\hspace{-15truept}
            \fig{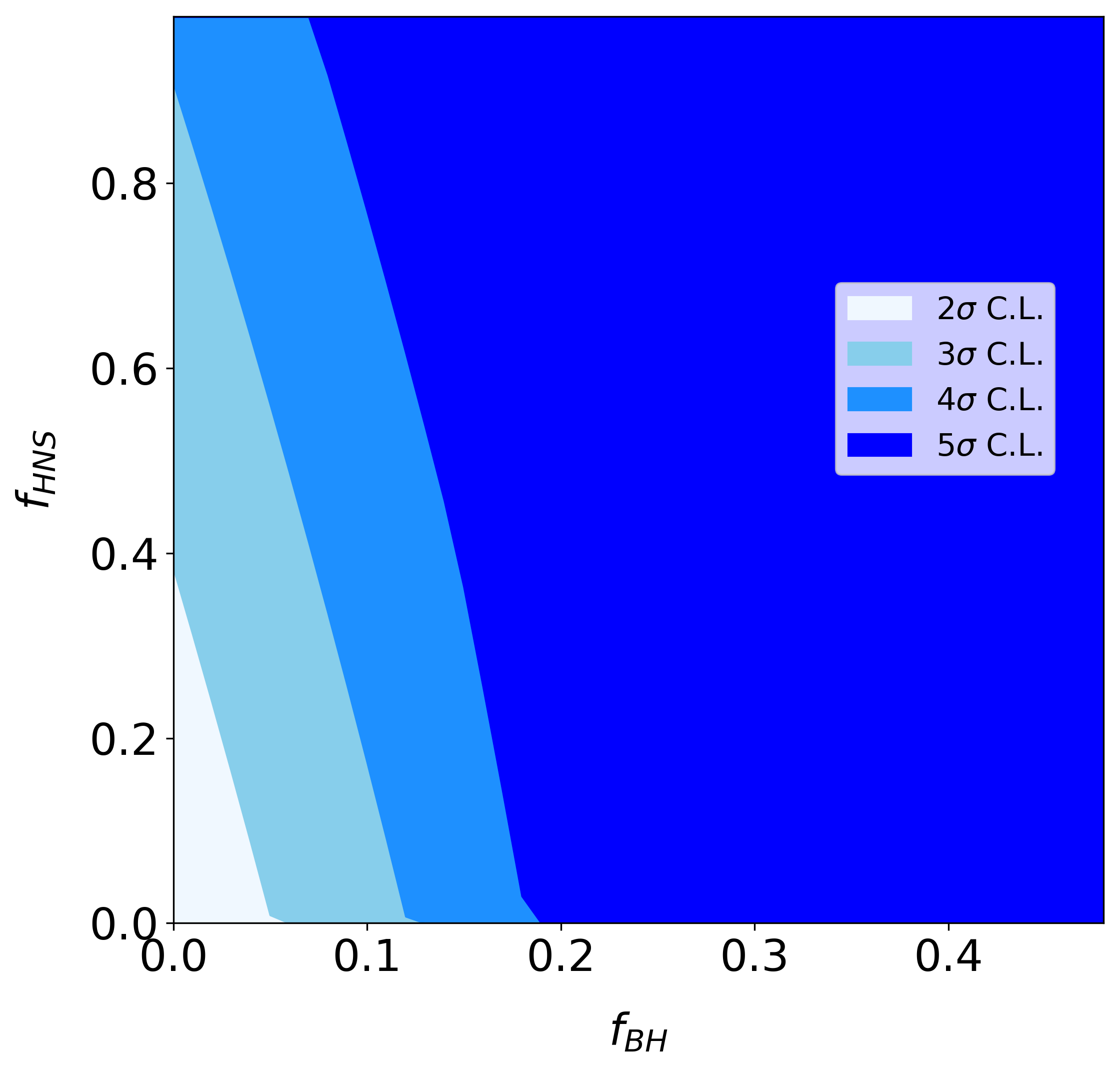}{0.32\textwidth}{(1c) Shen EOS, NH}}
  \gridline{\fig{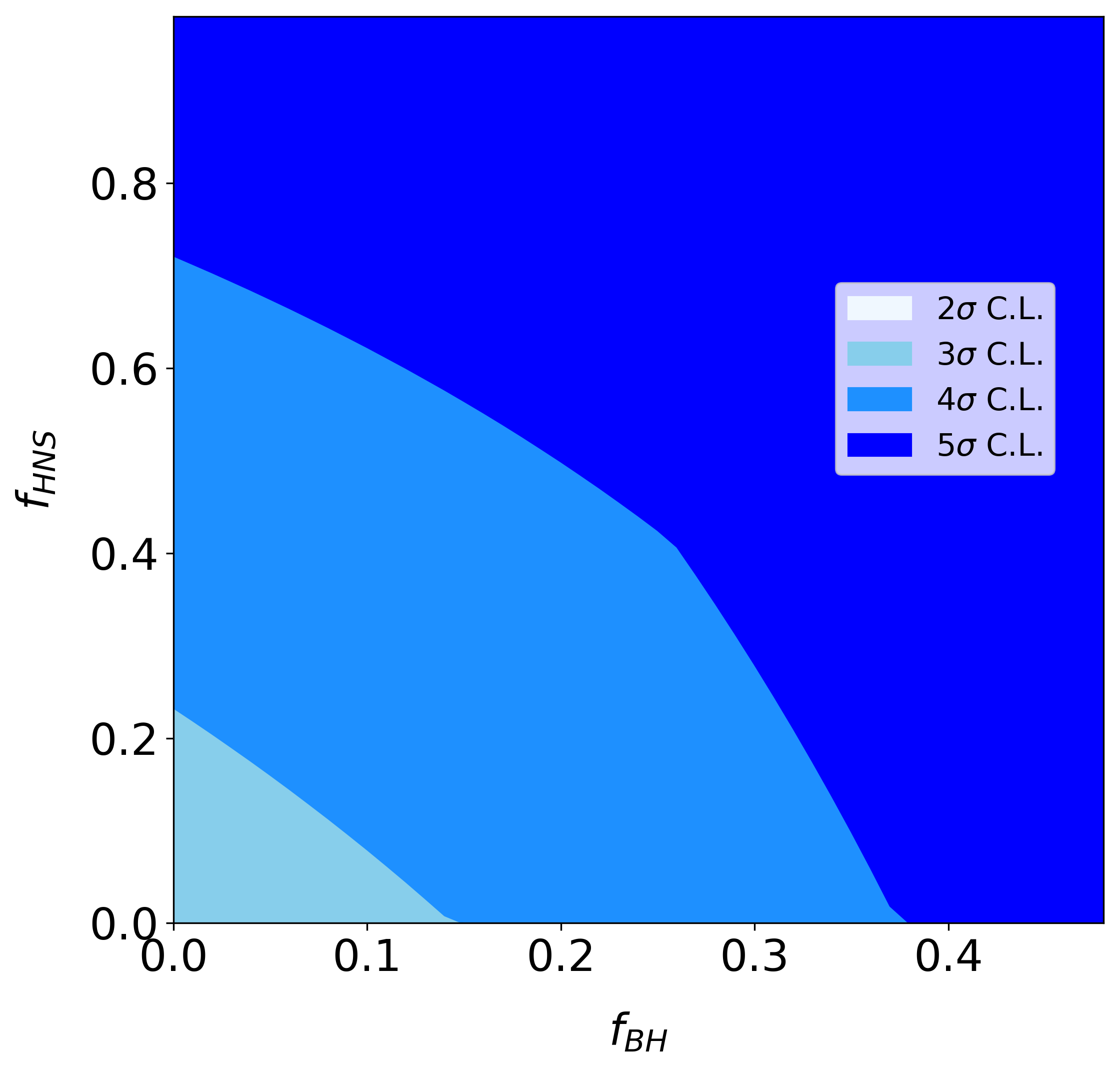}{0.32\textwidth}{(2a) Togashi EOS, IH}\hspace{-15truept}
            \fig{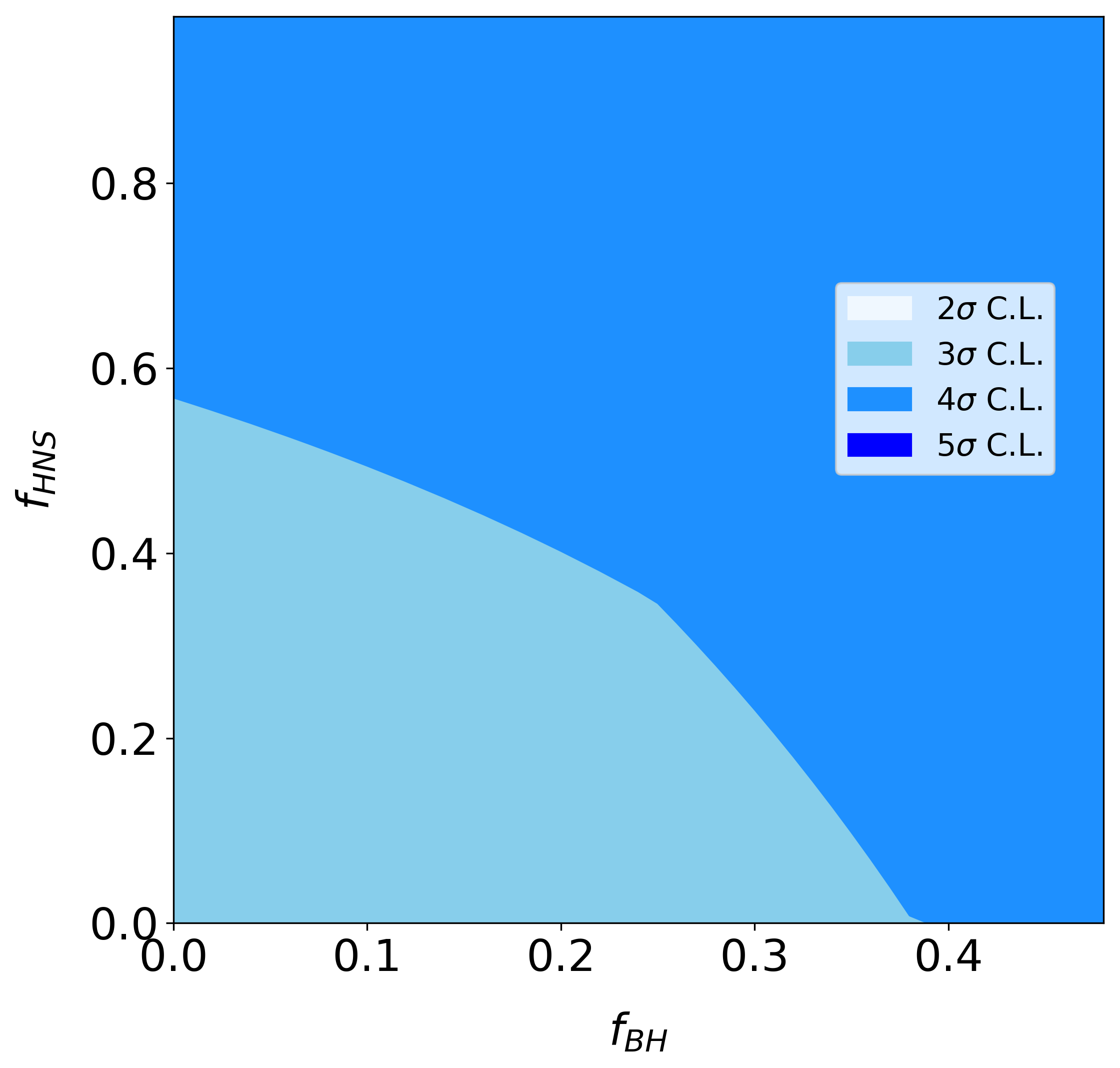}{0.32\textwidth}{(2b) LS220 EOS, IH}\hspace{-15truept}
            \fig{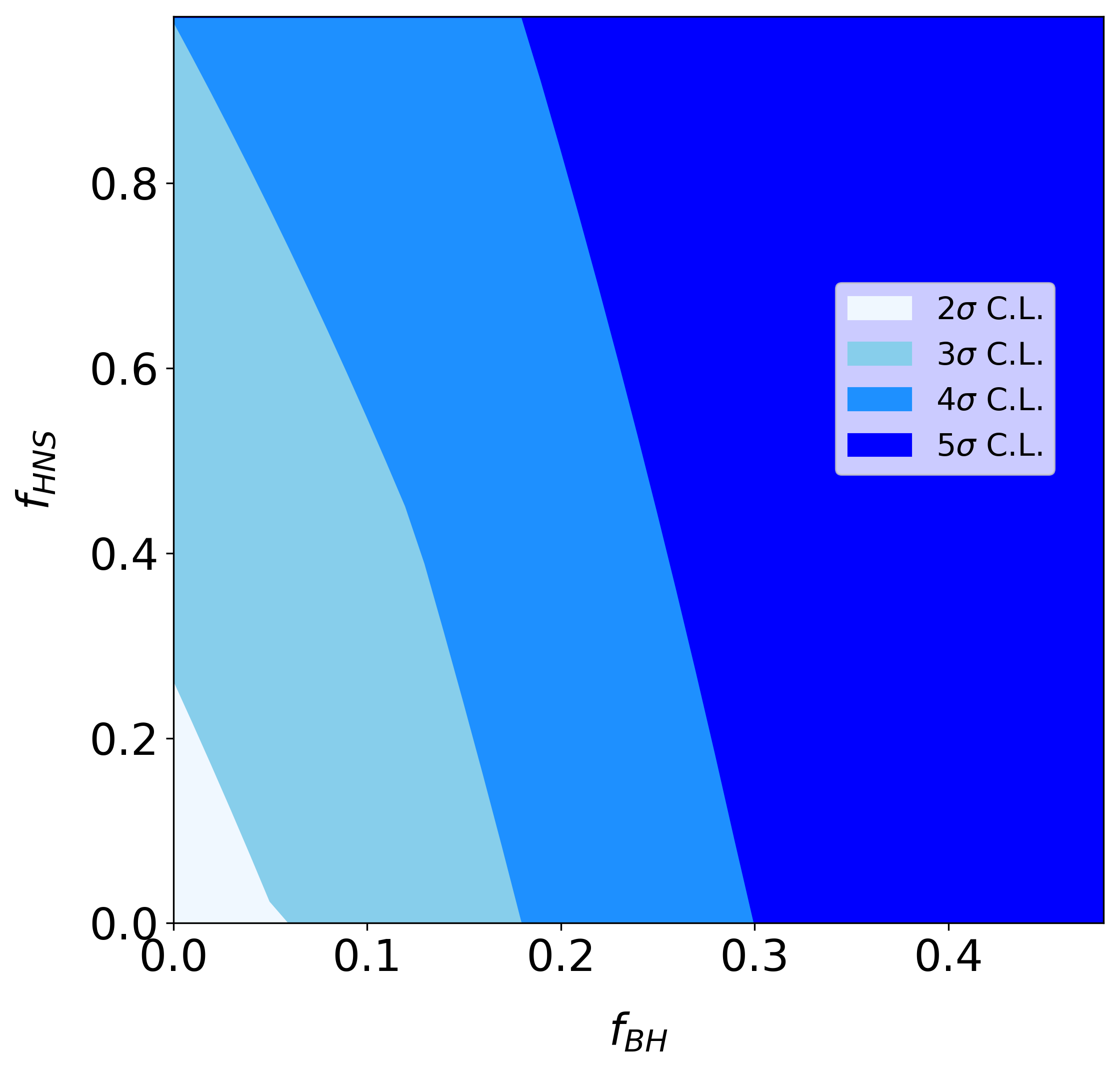}{0.32\textwidth}{(2c) Shen EOS, IH}}
  \vspace{+5truept}
  \caption{Same distributions as Figure~\ref{fig:cntr2dyear} but with a doubled amount of neutrino emission from the BH formation.}
  \label{fig:cntr2dyearbhsyst}
  \end{figure*}

The gravitational energy released in a form of neutrinos during the failed process strongly depends on the progenitor structure~\citep{O_Connor_2011,2018MNRAS.475.1363H,2020ApJ...894....4D,2021ApJ...909..169K}, thus constraints on the fractional parameters shown in Section~\ref{sec:snstv} may change for different models of the failed SN forming a BH.
In \citet{2021ApJ...909..169K}, the failed SN case that forms a BH with a significant delay ($\gg$1~sec) is studied in which about twice larger neutrino emission energy can be expected than the fast BH formation case.
We therefore double the neutrino emission amount of our BH model and study contours with this assumption to investigate systematic uncertainty for the delayed BH formation.
The results for HK over 10 yr are shown in Figure~\ref{fig:cntr2dyearbhsyst}.
In general, as expected, $f_{\rm BH}$ can be constrained more strictly for the increased neutrino amount, which is more obvious for the Shen EOS whose neutrino emission from BH is greater (see Figure~\ref{fig:neventcomp}). 

  %%%
  %%%

\begin{figure*}[htbp]
  \gridline{\fig{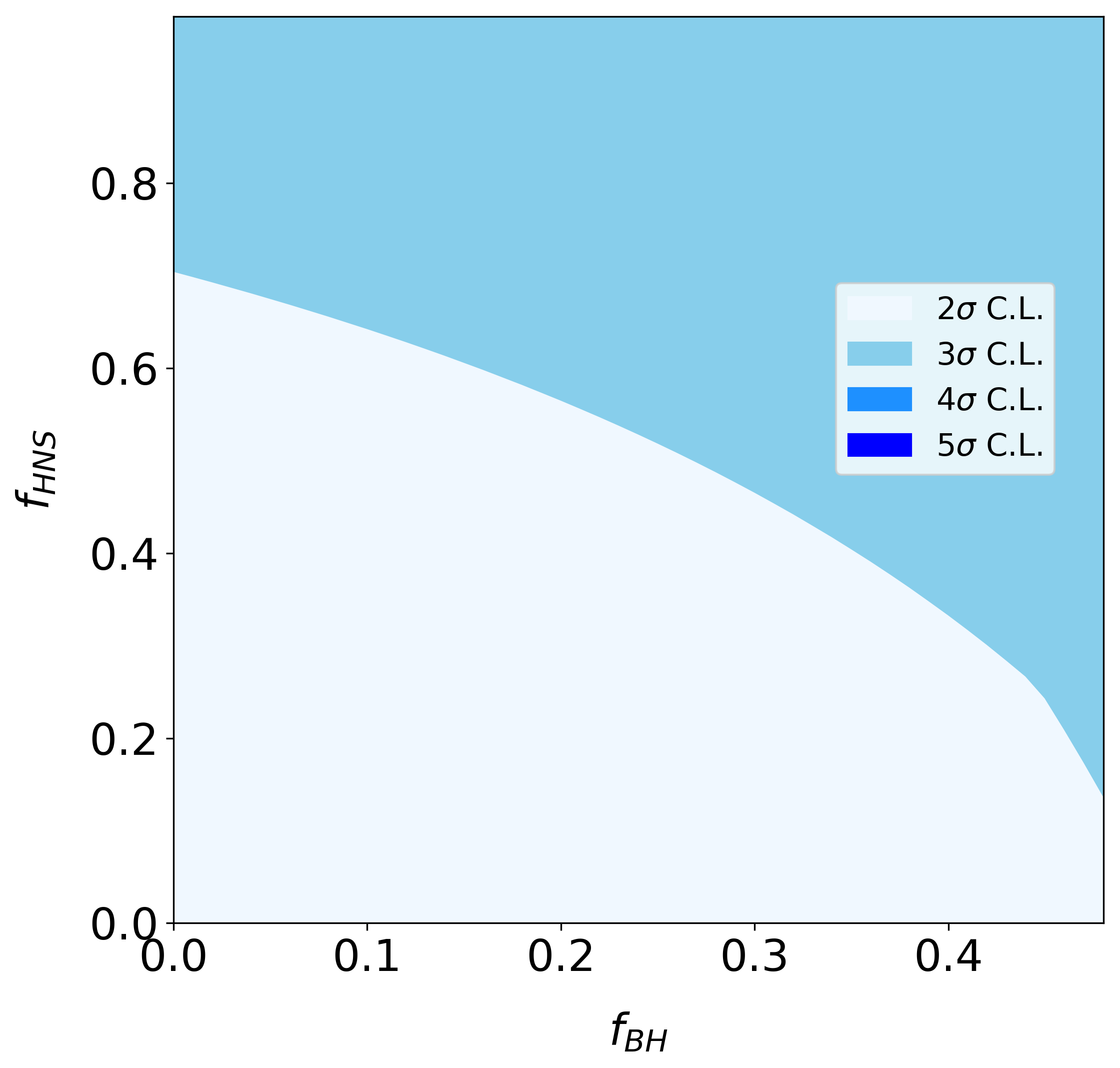}{0.32\textwidth}{(1a) Togashi EOS, NH}\hspace{-15truept}
            \fig{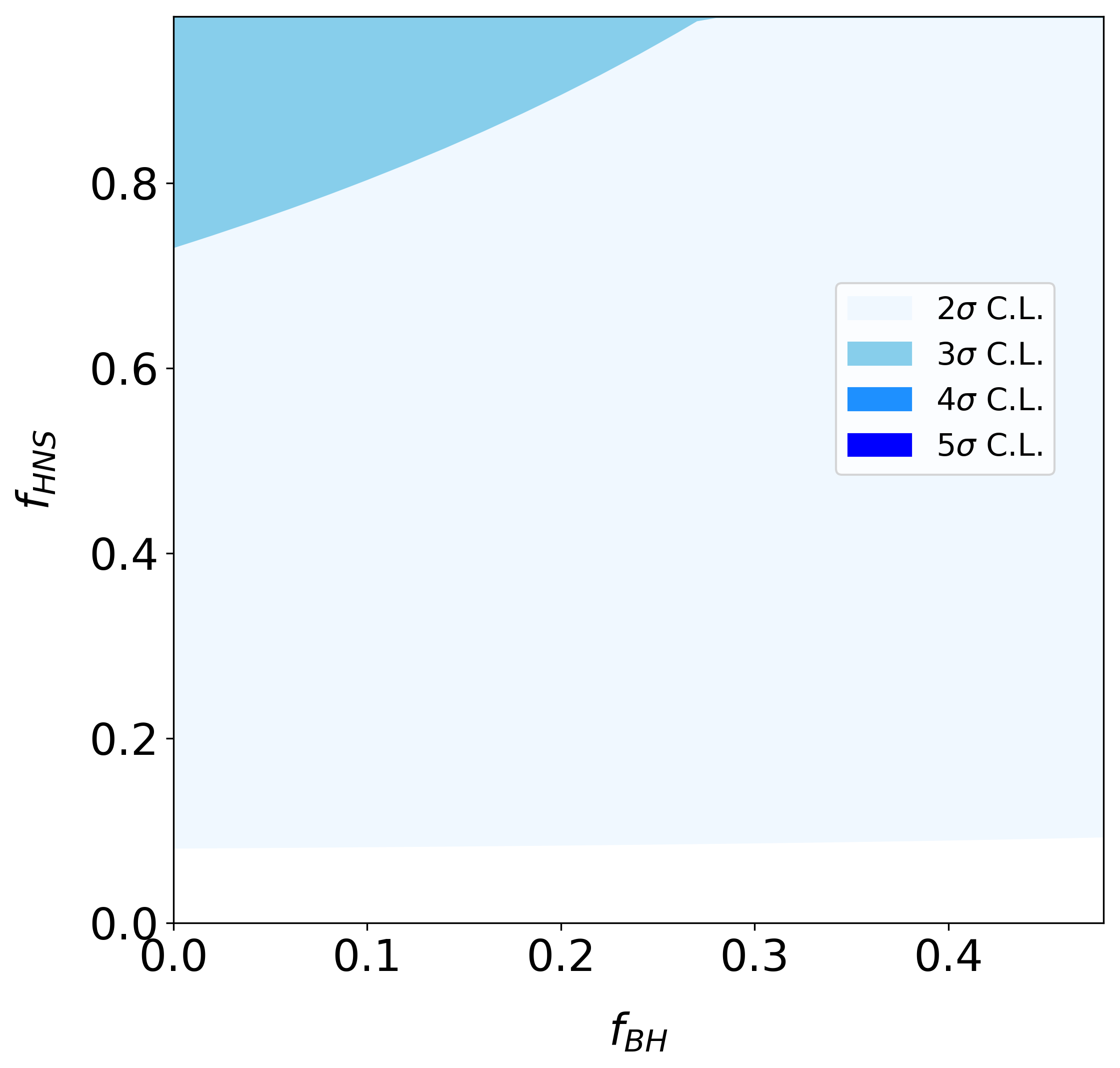}{0.32\textwidth}{(1b) LS220 EOS, NH}\hspace{-15truept}
            \fig{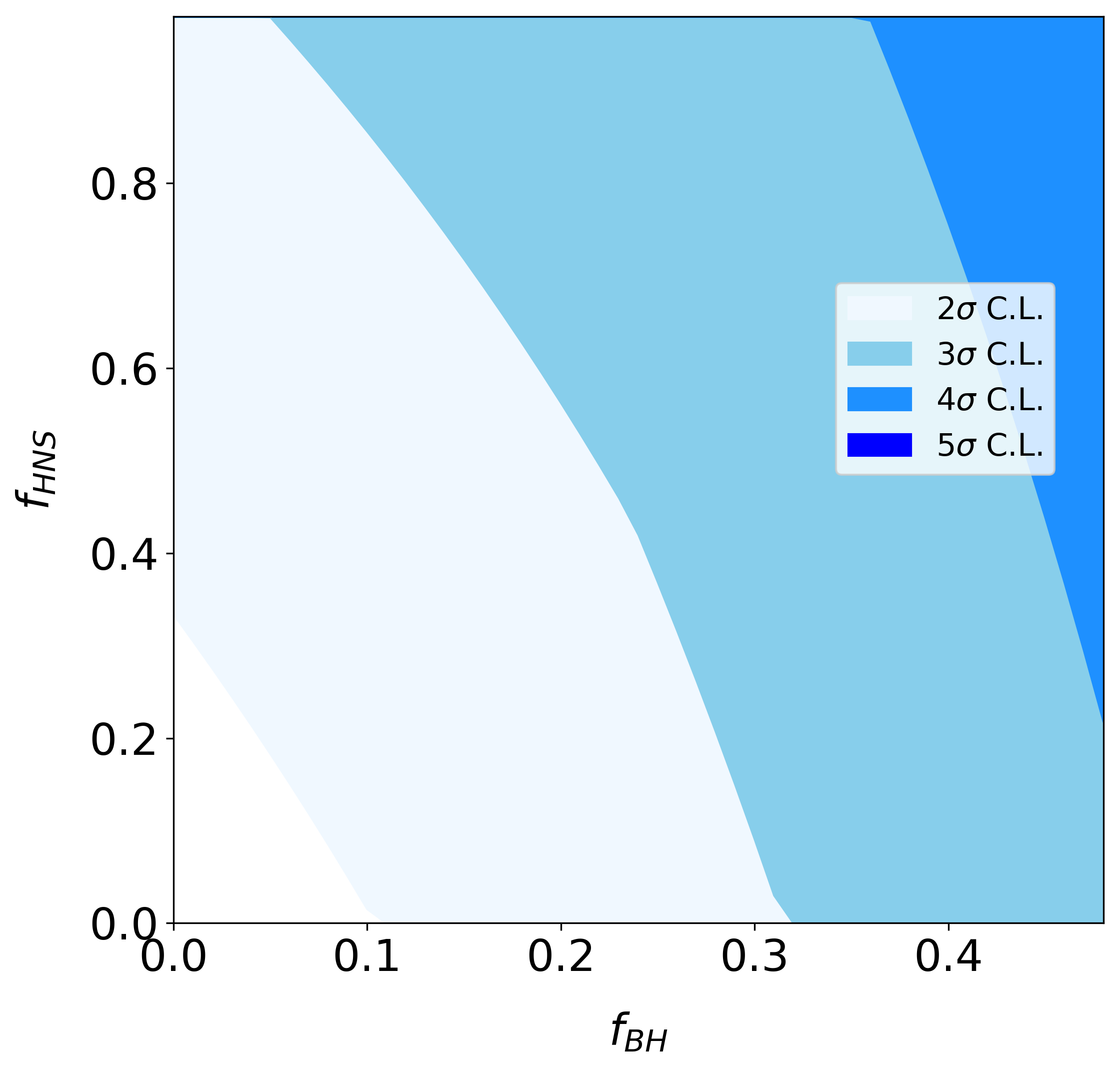}{0.32\textwidth}{(1c) Shen EOS, NH}}
  \gridline{\fig{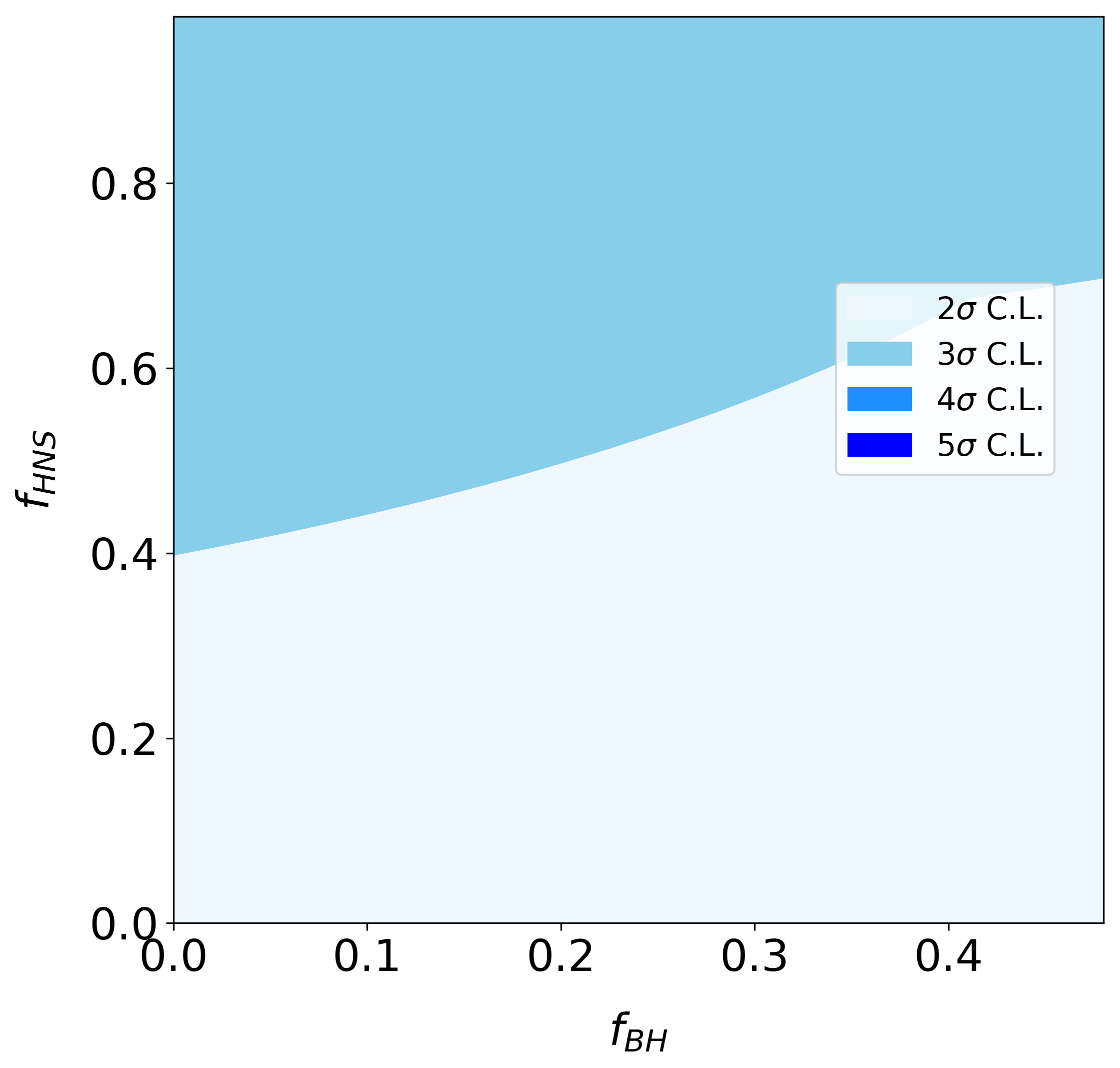}{0.32\textwidth}{(2a) Togashi EOS, IH}\hspace{-15truept}
            \fig{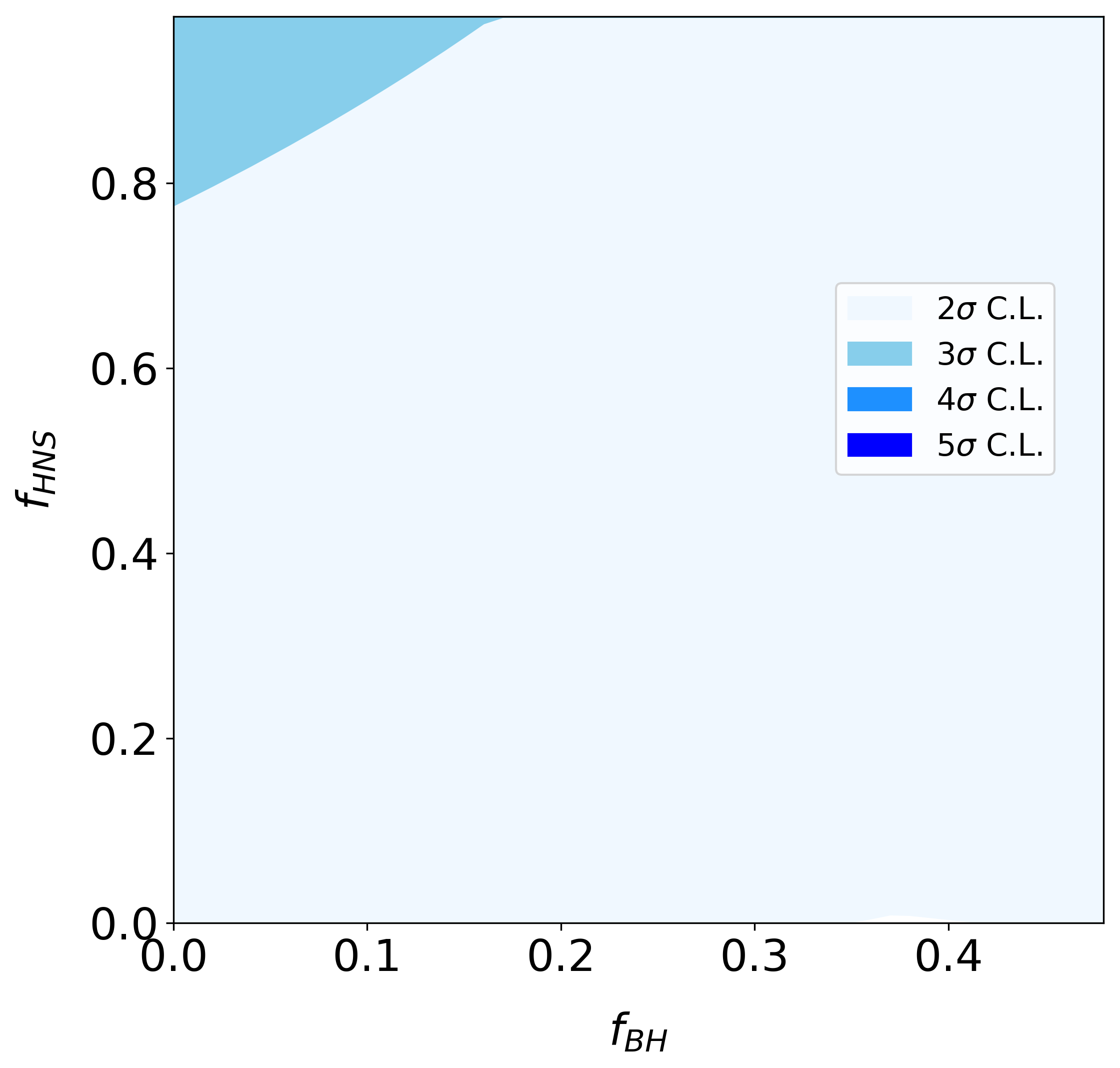}{0.32\textwidth}{(2b) LS220 EOS, IH}\hspace{-15truept}
            \fig{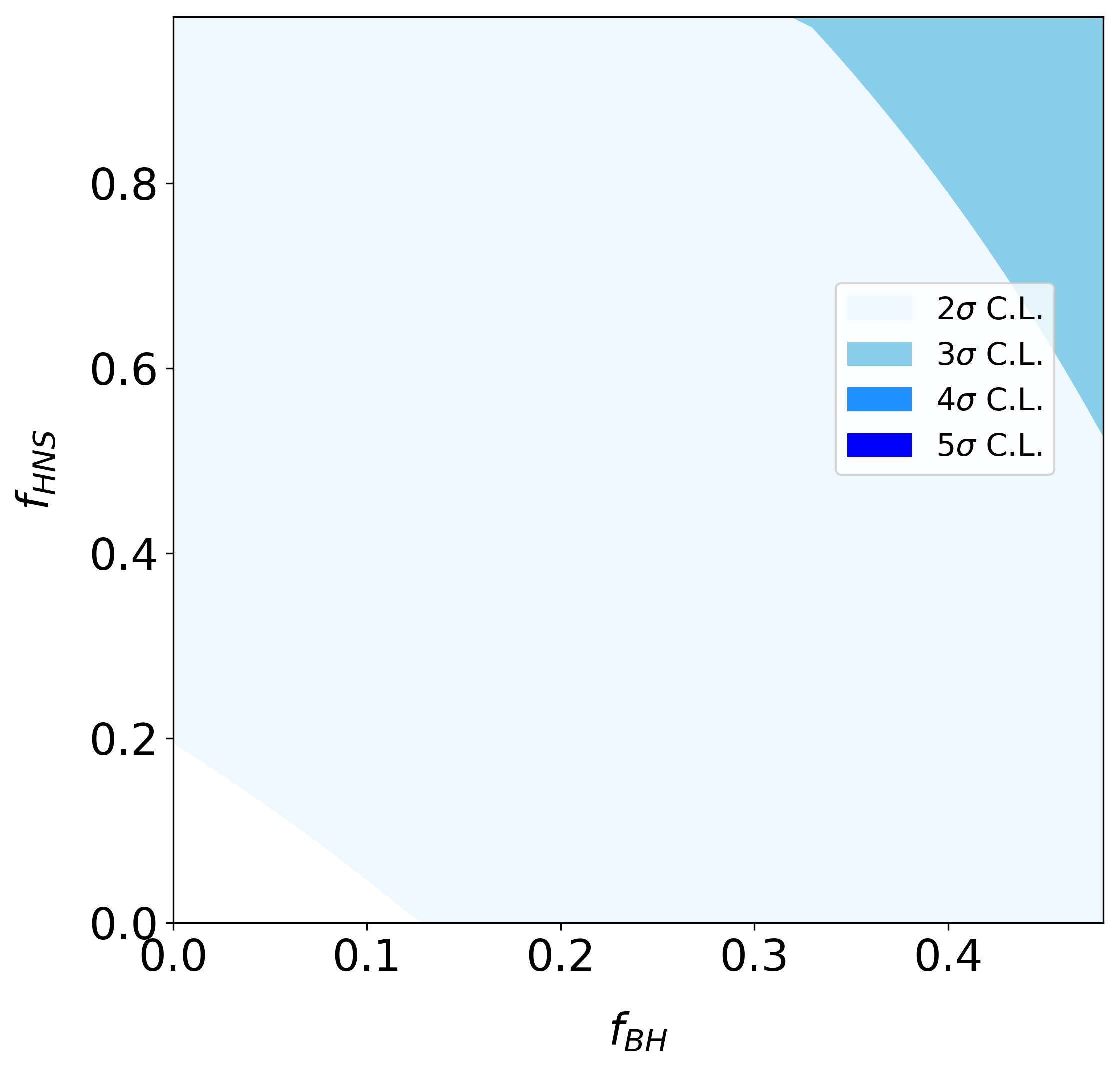}{0.32\textwidth}{(2c) Shen EOS, IH}}
  \vspace{+5truept}
  \caption{Same distributions as Figure~\ref{fig:cntr2dyear} but with $M_{\rm min}=8M_\odot$ in Equation~(\ref{eq:ccrate}).}
  \label{fig:cntr2dyearmminsyst}
  \end{figure*}

Our SRN flux calculation is based on the minimum mass of progenitors being $M_{\rm min}=6M_\odot$ as described in Section~\ref{sec:form}, while larger values are used in many other studies. 
In order to investigate the impact of $M_{\rm min}$ on the current study, we calculate the SRN flux with $M_{\rm min}=8M_\odot$ and draw contours on the fractional parameters in Figure~\ref{fig:cntr2dyearmminsyst} for the case of HK over 10 yr. 
The significance is lower for all cases because of the smaller number of progenitors that contribute to neutrino emission, while the large parameter space can still be tested at a $2\sigma$ level. 

  %%%
  %%%

We showed the SRN flux for the different EOS and neutrino mass hierarchy cases, while these could be constrained by external studies. The constraints on the EOS are provided by the mass measurements of heavy NSs \citep[][]{2013Sci...340..448A,2018ApJS..235...37A,2020NatAs...4...72C,2021ApJ...908L..46R}, the NS radius from the LIGO-Virgo detection of GW170817 \citep[][]{2018PhRvL.121p1101A}, and  the NS mass and radius from the NICER observation \citep[][]{2019ApJ...887L..21R,2019ApJ...887L..24M,2021ApJ...918L..28M}. As shown in Figure 1 of \citet{2022ApJ...925...98N}, the Togashi EOS is more favored than the other EOSs; the LS220 EOS has the maximum mass of cold NSs ($2.04M_\odot$) that is lower than the mass of heavy NSs already discovered and the NS radius of the Shen EOS is larger than the estimation from the gravitational wave data. Further details on the EOS and its constraints are seen in \citet{2022ApJ...925...98N}. NH is currently favored by the neutrino oscillation measurements (see \citet{10.1093/ptep/ptz015} for example), while the statistical significance is not enough to make a conclusion. Planned oscillation experiments are expected to confirm this with a higher confidence~\citep[][]{10.1093/ptep/pty044,EPJC80.978.2020,EPJC82.26.2022}. 

In the present study, we assume that the fraction of failed SNe ($f_{\rm BH}$) is independent of the redshift. However, it would not be the case because the fate of massive stars is determined by metallicity. The metallicity distribution function of progenitors at the selected redshift can also be derived from the data of Illustris-1 result \citep[][]{2015A&C....13...12N}, which is adopted for the cosmic star formation rate in our study, and the redshift dependence of $f_{\rm BH}$ will be a subject of future investigations.

Our sensitivity studies are based on realistic assumptions extrapolated from the SK results, but further efforts could enhance the detector's power to test the parameter space. For example, reducing the atmospheric NCQE background would improve the sensitivity dramatically at lower energies. This could be achieved by a more precise reconstruction of the neutron vertex. In addition, external data about neutrino interaction cross sections and isotope productions in a muon spallation would help to predict backgrounds more precisely. This becomes more important especially at larger volume detectors such as HK.

%%---------------------------------------------------------------------------------
\section{Conclusion} \label{sec:concl}

In this paper, we focused on SRNs as a probe to investigate the fate of stellar core collapse. We prepared core-collapse SN models for three different fates: the SN with a canonical-mass NS, the SN with a high-mass NS, and the failed SN with a BH, for three different nuclear EOSs: Togashi, LS220, and Shen. These models show unique shapes in the resulting neutrino spectrum. We introduced two parameters, $f_{\rm BH}$ (the fraction of failed SNe to the total number of stellar core-collapse events) and $f_{\rm HNS}$ (the fraction of SNe with a high-mass NS to the total number of SNe with an NS), to mix these three fate components to form the SRN flux. We performed the sensitivity study at the next-generation water-based Cherenkov detector, SK-Gd and HK, as extrapolating from the latest SK analysis to achieve realistic assumptions. As a result, we found that the sensitivity depends mainly on $f_{\rm HNS}$ if the delayed BH formation cases are neglected and the EOS has a low maximum mass of NSs as is the case for the LS220 EOS. In contrast, for the EOS with a higher maximum mass such as the Shen EOS, the impact of $f_{\rm BH}$ gets stronger. We also showed that the analysis in different energy ranges would constrain different areas in a $f_{\rm BH}$--$f_{\rm HNS}$ space. The sensitivity with HK over 10 yr shows that most of the area in this space could be tested at a $3\sigma$ level. The results of this paper help to provide implications about the fate of stellar core collapse in compound with other studies such as EOS and neutrino mass hierarchy.

\begin{acknowledgments}
This work is supported by Grant-in-Aid for Scientific Research (JP20K03973) and Grant-in-Aid for Scientific Research on Innovative Areas (JP19H05811) from the Ministry of Education, Culture, Sports, Science and Technology (MEXT), Japan.
\end{acknowledgments}

%%---------------------------------------------------------------------------------
%\appendix

%\section{}

%%---------------------------------------------------------------------------------
%% For this sample we use BibTeX plus aasjournals.bst to generate the
%% the bibliography. The sample631.bib file was populated from ADS. To
%% get the citations to show in the compiled file do the following:
%%
%% pdflatex sample631.tex
%% bibtext sample631
%% pdflatex sample631.tex
%% pdflatex sample631.tex
%\clearpage
\bibliography{sample631}{}
\bibliographystyle{aasjournal}

%% This command is needed to show the entire author+affiliation list when
%% the collaboration and author truncation commands are used.  It has to
%% go at the end of the manuscript.
%\allauthors

%% Include this line if you are using the \added, \replaced, \deleted
%% commands to see a summary list of all changes at the end of the article.
%\listofchanges

\end{document}